\documentclass{article}

\usepackage{PRIMEarxiv}

\usepackage[utf8]{inputenc} 
\usepackage[T1]{fontenc}    
\usepackage{hyperref}       
\usepackage{url}            
\usepackage{booktabs}       
\usepackage{amsfonts}       
\usepackage{nicefrac}       
\usepackage{microtype}      
\usepackage{lipsum}
\usepackage{graphicx}
\graphicspath{{media/}}     
\usepackage{subcaption}
\usepackage{verbatim}
\usepackage{float}
\usepackage{amsmath}
  
\title{

Where Do Your Citations Come From? Citation-Constellation: A Free, Open-Source, No-Code, and Auditable Tool for Citation Network Decomposition with Complementary BARON and HEROCON Scores
\thanks{\textit{The acronyms BARON and HEROCON serve as metaphors that encode the conceptual relationship between the scores, as detailed in Section \ref{sec:metaphor_meaning}}.}}

\author{
  Mahbub Ul Alam \\
 SciLifeLab Data Centre, Uppsala University, Sweden \\
  \texttt{mahbub.ul.alam@scilifelab.uu.se, mahbub.ul.alam.anondo@gmail.com} \\
}

\begin{document}
\maketitle

\begin{abstract}

\begin{figure}
   \centering
    \includegraphics[scale=.15]{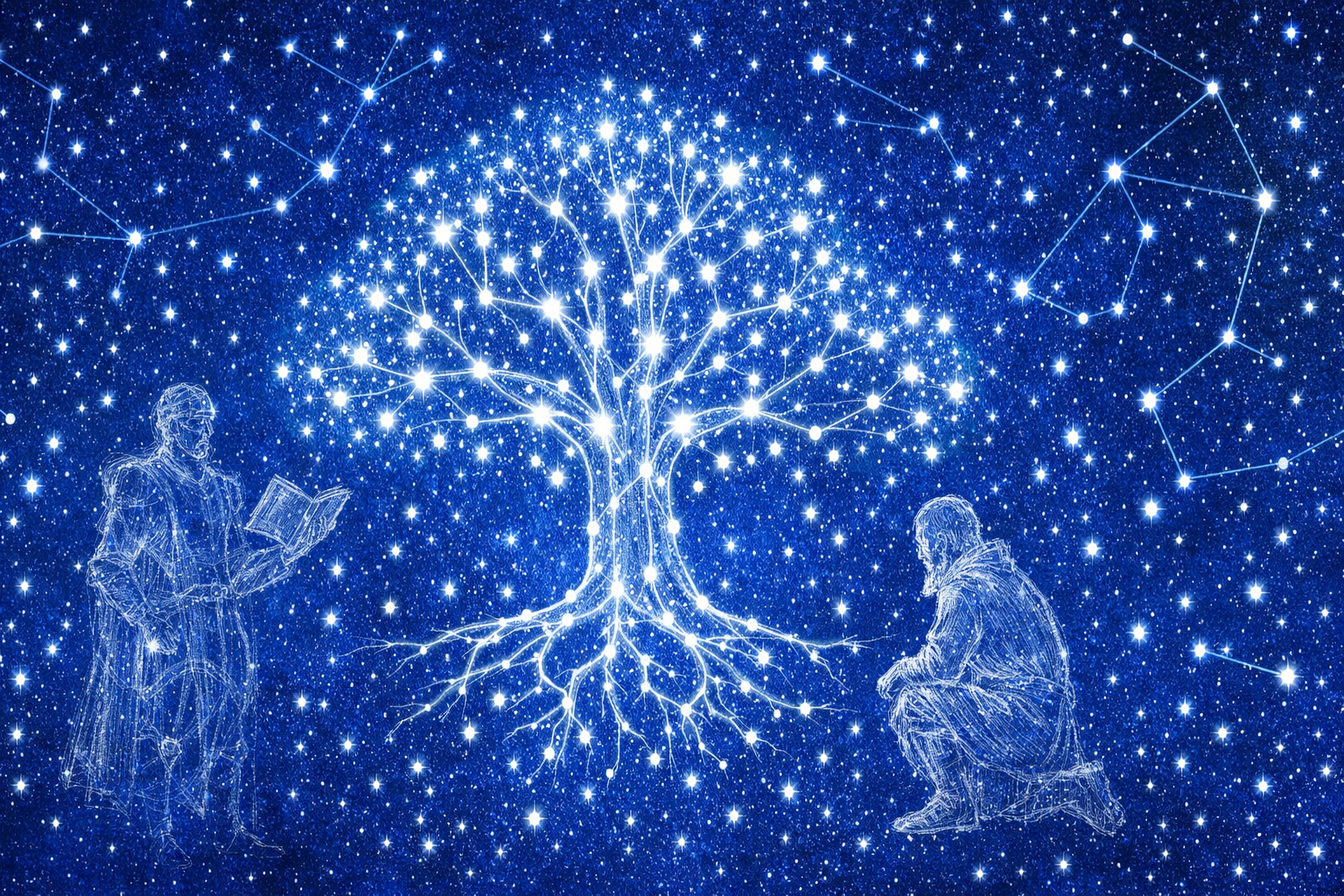}
    \label{cover}
\end{figure}

In this paper, I introduce Citation-Constellation, a fully operational and freely available tool\footnote{\textbf{Citation-Constellation No-Code Tool:} \href{https://citation-constellation.serve.scilifelab.se}{citation-constellation.serve.scilifelab.se}} for citation network analysis. Every claim presented here can be verified directly by the reader, with no installation, registration, or payment required. Users may enter an ORCID or OpenAlex ID to receive a complete citation network decomposition within minutes. The source code is publicly available\footnote{\textbf{Citation-Constellation Source Code:} \href{https://github.com/citation-cosmograph/citation-constellation}{github.com/citation-cosmograph/citation-constellation}}.

Standard citation metrics treat all citations as equal, thereby obscuring the social and structural pathways through which scholarly influence actually propagates \cite{waltman2016review,hicks2015bibliometrics}. A citation from an independent researcher who discovers a paper through literature search carries a different epistemic weight than one from a direct co-author, a departmental colleague, or an editorial board member \cite{ioannidis2015generalized,  wallace2012small}. Yet conventional metrics such as the h-index, raw citation counts, and journal impact factors collapse these important distinctions into a single number.

I introduce two complementary bibliometric scores that decompose a researcher's citation profile according to network proximity between citing and cited authors. BARON (Boundary-Anchored Research Outreach Network score) provides a strict binary metric that counts only citations from outside the detected collaborative network, while HEROCON (Holistic Equilibrated Research Outreach CONstellation score) applies configurable graduated weights that assign partial credit to in-group citations based on relationship proximity. The gap between these two scores serves as a diagnostic indicator of inner-circle dependence.

The accompanying open-source tool implements this decomposition through a phased detection architecture comprising (1) self-citation analysis, (2) co-authorship graph traversal, (3) temporal institutional affiliation matching via the Research Organization Registry (ROR), and (4) AI-agent-driven venue governance extraction using a locally deployed large language model\footnote{\textbf{Citation-Astrolabe Source Code (venue governance database, under construction):} \href{https://github.com/citation-cosmograph/citation-astrolabe}{citation-cosmograph/citation-astrolabe}}. The first three phases are fully operational and accessible via the provided web tool, while Phase 4 is currently under active implementation. The implementation places particular emphasis on ORCID-validated author identity resolution, an UNKNOWN  classification for citations with insufficient metadata, and comprehensive audit trails that document every classification decision. The venue governance phase demonstrates how recent advances in locally deployable generative AI can automate bibliometric infrastructure tasks, such as editorial board extraction and entity resolution, that were previously intractable without dedicated research teams. A no-code web interface deployed on SciLifeLab Serve enables researchers to compute and visualize their scores without programming knowledge, installation, or registration.

I present these scores as structural diagnostics that describe \textit{where} in the social graph citations originate, explicitly not as quality indicators for research evaluation, and discuss their alignment with the responsible research assessment movement \cite{dora2013, hicks2015bibliometrics}. The graduated HEROCON weights remain experimental and require empirical calibration; I identify sensitivity analysis and cross-validation with citation motivation studies as priorities for future work.

\end{abstract}

\keywords{bibliometrics \and citation analysis \and citation network structure \and co-authorship network \and self-citation \and author disambiguation \and research impact measurement \and network proximity \and responsible research assessment \and open science \and research evaluation \and scientometrics \and scholarly communication \and venue governance \and institutional affiliation \and ORCID \and audit trail \and reproducibility \and large language model \and generative AI \and AI agent \and AI infrastructure \and automated information extraction \and natural language processing}

\section{Introduction}\label{sec:introduction}

The quantification of research impact represents a central concern of modern science policy, institutional evaluation, and individual career assessment. Metrics such as the h-index \cite{hirsch2005index}, citation count, and journal impact factor have become deeply embedded in academic hiring, promotion, and funding decisions. Yet these metrics share a fundamental limitation: they treat all citations as equivalent signals of impact, regardless of the relationship between the citing and cited researchers. This limitation has been noted by numerous scholars \cite{seglen1997impact, bornmann2005does, costas2007h, hicks2015bibliometrics, waltman2016review}.

A citation from an independent researcher who discovers a paper through literature search carries a different epistemic weight than a citation from a direct co-author, a colleague in the same department, or one that arrives through a venue where the cited researcher holds an editorial or governance role. The former represents genuine external reach, providing evidence that the work has found relevance beyond its originating community. The latter, while not illegitimate, reflects the natural amplification effects of collaborative networks, institutional proximity, and professional relationships. This phenomenon, the tendency for citation patterns to follow social network structure, has been extensively studied in the literature on invisible colleges \cite{crane1972invisible, zuccala2006modeling}, co-authorship networks \cite{newman2001structure, newman2004coauthorship, glanzel2004analysing, moody2004structure}, and citation homophily \cite{wallace2012small, wagner2005network}.

The core insight, that citation patterns are shaped by social network proximity, is well-established in network science and scientometrics. However, operationalizing this insight into a practical, auditable tool has been hindered by persistent challenges in author disambiguation, metadata quality, and the complexity of building comprehensive network detection layers that span self-citation, co-authorship, institutional affiliation, and venue governance. I address these challenges with an open-source tool that makes several key engineering contributions.

First, I present Citation-Constellation, a phased detection architecture that progressively deepens network analysis from self-citation through co-authorship graphs, temporal affiliation matching, and venue governance detection, with each phase producing an independently meaningful and usable score.

Second, the tool incorporates ORCID-validated author identity resolution to address the critical challenge of disambiguation error contaminating citation analysis.

Third, an UNKNOWN classification honestly reports citations with insufficient metadata rather than silently defaulting them to EXTERNAL, preventing systematic bias against researchers with poor metadata coverage.

Fourth, comprehensive audit transparency ensures that every classification decision is documented with a human-readable rationale in a structured JSON file, enabling full reproducibility and contestability.

Fifth, the Phase 4 pipeline employs an AI-agent-driven venue governance extraction system using a locally deployed language model to build a persistent, incrementally expanding database of editorial board and program committee membership across disciplines.

Finally, the dual-score framework, comprising BARON (Boundary-Anchored Research Outreach Network score, a strict binary metric) and HEROCON (Holistic Equilibrated Research Outreach CONstellation score, a graduated weighted metric), provides an interpretable summary in which the gap between scores serves as a structural diagnostic of inner-circle dependence.

To ensure full accessibility for researchers regardless of their technical background, Citation-Constellation is available as both a command-line application and a no-code web interface (see Table \ref{tab:citation-cosmograph-ecosystem} for links).

I emphasize that the HEROCON weights are experimental defaults informed by reasoning about relationship proximity, not empirically calibrated values. I present these tools as structural diagnostics, not quality indicators. BARON and HEROCON describe where in the social graph citations originate. They should not be used for hiring, promotion, or funding decisions.

\section{Positioning My Contribution}\label{sec:my_position}

The core insight, that citation patterns reflect network structure, is well-established, from Crane's invisible colleges \cite{crane1972invisible} through Wallace et al.'s network-distance analysis \cite{wallace2012small} and beyond. My contribution lies in synthesizing these disparate lines of inquiry into a unified, operational framework. By combining co-authorship analysis \cite{newman2001structure, newman2004coauthorship}, institutional proximity detection \cite{lariviere2010impact, pan2012world}, and editorial governance connections (novel in this context) with robust ORCID-based identity resolution \cite{haak2012orcid}, honest data quality reporting, and full audit transparency, I transform a theoretical lens into a practical diagnostic instrument.

This is not merely engineering. The history of scientometrics shows that the gap between theoretical insight and practical deployment is where most value is lost \cite{wildgaard2014review, waltman2016review}, and that this gap is not neutral. When powerful analytical tools exist only as conceptual frameworks in journal articles, they benefit only those with the technical resources to reimplement them \cite{ince2012case}. When they exist as accessible, free, open-source software, they become democratic infrastructure \cite{priem2022openalex}.

\subsection{Accessibility as a Contribution}

I deploy the Citation-Constellation tool as a web application on SciLifeLab Serve (\href{https://citation-constellation.serve.scilifelab.se}{citation-constellation.serve.scilifelab.se}), freely accessible to anyone with a web browser. It requires no installation, no programming knowledge, no institutional subscription, no registration, and no payment. A researcher may navigate to the URL, enter an ORCID or OpenAlex identifier, and receive a full multi-layer citation network decomposition with interactive visualizations and a downloadable audit report within minutes.

This matters because many researchers understand, in principle, that citation patterns reflect network structure. Far fewer have access to a tool that decomposes their own profile across multiple network layers, validates author identity, honestly reports data limitations, and documents every decision in a contestable audit trail. Bibliometric self-knowledge has historically been a privilege of researchers with access to expensive proprietary databases such as Scopus and Web of Science \cite{martin2021google, visser2021large} and the technical skills to query them. A researcher at a well-funded university and a researcher at a teaching-intensive institution in the Global South face very different barriers to understanding their own citation structure, barriers that are economic and technical, not intellectual \cite{tennant2020web, demeter2020academic}. By building entirely on open data sources, OpenAlex \cite{priem2022openalex}, ORCID \cite{haak2012orcid}, and ROR \cite{lammey2020solutions}, and by providing both a command-line tool and a no-code web interface, I lower the barrier to citation profile analysis to near zero.

The comparison feature, which allows side-by-side visualization of multiple researcher profiles from uploaded audit files, further enables peer learning: colleagues within a department, a field, or a cohort may examine their structural patterns together, generating insights about disciplinary norms, collaboration effects, and career trajectories that no individual analysis could reveal. A department head could compare the citation constellations of their team to understand structural differences; a doctoral student could benchmark their emerging profile against established researchers in their field. These are not evaluative comparisons. They are structural explorations, made possible only when the tool is freely and universally accessible. This aligns with the broader open science movement's emphasis on making research infrastructure equitable and transparent \cite{unesco2021openscience, peroni2020opencitations, waltman2024openness, barcelona2024}.

\subsection{The Audit Trail as Curated Evidence}

The BARON/HEROCON dual-score framework provides an interpretable summary, but I consider the audit trail to be the deeper contribution. Each audit file is a structured, timestamped, machine-readable record of every citation link and every classification decision for a specific researcher at a specific moment in time. This is not merely a transparency mechanism. It is a curated data product. Aggregated across researchers, fields, and institutions, these audit files constitute a novel dataset for science-of-science research \cite{fortunato2018science}: one that captures not merely \textit{how many} citations a researcher received, but the network relationship underlying each one.

The hardest barrier in bibliometric research has always been the curation of trustworthy, structured data sources linking citations to their social context \cite{waltman2016review, huang2020comparison, visser2021large}. Existing large-scale citation databases \cite{ioannidis2019standardized, priem2022openalex} capture citation counts and field classifications but do not decompose individual citations by the network relationship between citing and cited authors. My audit trail addresses this gap directly: each file is self-contained, reproducible, and carries its own provenance metadata. Future analyses, whether on collaboration norms across fields, on the evolution of collaboration networks and productivity \cite{petersen2012persistence}, citation networks over career stages \cite{borner2004simultaneous, yin2017time}, or on institutional effects on citation composition \cite{lariviere2010impact}, can build on this curated foundation without repeating the expensive data collection and classification pipeline.

My emphasis on auditability also responds to growing calls for transparency in research evaluation \cite{wilsdon2015metrictide, rijcke2016evaluation}. Traditional metrics are opaque: a researcher receives an h-index but cannot inspect how each citation was counted or whether misattributed works inflated the number. The Citation-Constellation audit trail, by contrast, makes every assumption contestable.

\subsection{AI-Agent-Driven Bibliometric Infrastructure}

Several components of this tool, particularly the venue governance extraction in Phase 4, would have been considered impractically labor-intensive even a year ago. Manually identifying editorial board members and program committee members across hundreds of academic venues, resolving their identities against bibliometric databases, and maintaining a persistent, incrementally growing governance database would require a dedicated research team. The rapid maturation of locally deployable large language models \cite{touvron2023llama, bai2023qwen} and efficient inference engines such as llama.cpp \cite{gerganov2023llamacpp} has fundamentally changed this calculus. A quantized 8-billion-parameter model running on commodity CPU hardware can now perform structured information extraction from heterogeneous HTML pages with sufficient accuracy for entity resolution against bibliometric databases, a task that previously required either expensive commercial APIs, brittle rule-based scrapers, or dedicated human annotation \cite{ferreira2012brief}.

The resulting venue governance database is not merely useful for BARON/HEROCON scoring. It is an independent scholarly resource. It could support studies of editorial board diversity across disciplines \cite{mauleon2013assessing, palser2022gender}, analyses of the relationship between editorial roles and citation patterns \cite{baccini2019citation}, investigations of how governance structures evolve over time, or simply serve as a lookup tool for researchers curious about who governs the venues they publish in.

More broadly, I consider this work a proof of concept for AI-agent-driven bibliometric infrastructure. The combination of web scraping, LLM-based structured extraction, entity resolution against open bibliometric databases, and persistent incremental storage represents a reusable architectural pattern. The same pipeline that extracts editorial boards could be adapted to extract funding acknowledgements, author contribution statements, or conflict-of-interest declarations. This is all structured data that exists on journal websites but is not systematically captured or reliably structured in existing commercial databases \cite{alvarez2017funding, hendricks2020crossref}.

\section{How to Use Citation-Constellation}\label{sec:how_to_use}

\graphicspath{{media/}}

The tool is available in two forms: a \textbf{no-code web interface} (recommended for most users) and a \textbf{command-line interface} (for advanced users and automation). Both produce identical scores and audit trails.

\subsection{No-Code Web Interface}

The web interface is deployed on SciLifeLab Serve and requires no installation, no programming knowledge, and no account creation. Navigate to the application URL and use one of six tabs:

\begin{enumerate}
    \item \textbf{Run Analysis} --- Compute new BARON \& HEROCON scores.
    \item \textbf{View Existing Audits} --- Upload audit JSON files for visualization and comparison.
    \item \textbf{How to Run Here \& Install Locally} --- Step-by-step instructions.
    \item \textbf{How BARON \& HEROCON Work} --- Plain-language methodology.
    \item \textbf{Future Features} --- Planned Phase 4, 5, and 6 capabilities.
    \item \textbf{Full Research Paper} --- This paper, embedded for reference.
\end{enumerate}

\subsubsection{Running a New Analysis}\label{sec:hercon-weight-customization-1}

Navigate to the \textbf{Run Analysis} tab.

\begin{enumerate}
    \item \textbf{Enter a researcher identifier.} Provide an ORCID (e.g., \texttt{0000-0000-0000-0000}) or an OpenAlex ID (e.g., \texttt{A0000000000}). Full URLs are accepted; the tool extracts the ID automatically. ORCID format validation includes ISO 7064 Mod 11,2 checksum verification; invalid checksums produce a clear error message.

    \item \textbf{Optional: Set a career start year (default: 2000).} If OpenAlex has merged works from a different researcher with a similar name, enter the year your career began. All earlier works will be excluded. This is often the simplest fix for name collision.

    \item \textbf{Optional: Adjust co-author graph depth (default: 2).} Depth 1 counts only direct co-authors as in-group. Depth 2 (recommended) includes co-authors of co-authors. Depth 3 extends to three hops.

    \item \textbf{Optional: Enable manual validation.} Check ``Wait for my validation before discarding flagged papers'' to review ORCID-flagged papers as checkboxes before scoring proceeds.

    \item \textbf{Optional: Upload custom HEROCON weights.} Under the ``Advanced'' accordion, upload a JSON file to override default weights. Unspecified classifications use defaults.

    \item \textbf{Click ``Run Analysis.''} For a researcher with approximately 50--100 publications, expect 1--4 minutes. A progress message is displayed throughout.
\end{enumerate}

\subsubsection{Understanding the Results}

\begin{itemize}
    \item \textbf{Score Summary} --- BARON, HEROCON, gap, total citations, classifiable citations, reliability rating.
    \item \textbf{Classification Donut Chart} --- Proportional breakdown of citation origins with scores in the center.
    \item \textbf{Classification Summary Table} --- Each category with count, percentage, and HEROCON weight.
    \item \textbf{Co-Author Network Graph} --- Interactive force-directed network. Target researcher as gold node; direct co-authors in crimson (sized by shared papers); transitive co-authors in blue. Hover for details. Networks above 150 nodes are automatically pruned.
    \item \textbf{Career Trajectory Chart} --- Dual-line BARON and HEROCON over time, shaded gap area, cumulative citation bars.
    \item \textbf{Full Classification Table} --- Every citation with classification, confidence, phase, and human-readable rationale. Collapsible, with ``Export Citations as JSON'' option.
    \item \textbf{Interpreting the BARON--HEROCON gap} --- A small gap (below 3\%) indicates overwhelmingly external impact. A moderate gap (3--10\%) suggests meaningful inner-circle contribution alongside dominant external reach. A large gap (above 10\%) indicates that a significant portion of measured impact comes from the collaborative network. This is neither good nor bad; it characterizes citation structure, not quality.
    \item \textbf{Download Audit Report} --- Complete JSON audit for offline analysis or re-visualization.
\end{itemize}

\subsubsection{Comparing Multiple Researchers}

In the \textbf{View Existing Audits} tab, upload previously generated audit JSON files. A single file produces full visualization. Multiple files produce a comparison table, separate overlaid BARON and HEROCON trajectory charts, and expandable individual reports. Up to 115 simultaneous comparisons are supported. Files are validated against the expected schema; invalid files are rejected with clear messages.

\textbf{Rate limit:} 10 analyses per hour per session. Visualization has no limit.

\subsection{Command-Line Interface}

\subsubsection{Installation}

\begin{verbatim}
git clone https://github.com/citation-cosmograph/citation-constellation.git
cd citation-constellation
pip install -r requirements.txt
\end{verbatim}

Requirements: Python 3.11+. No database needed for Phases 1--3.

\subsubsection{Basic Usage}

\begin{verbatim}
# Phase 1: Self-citation baseline only
python phase1.py --orcid 0000-0000-0000-0000

# Phase 2: Self-citation + co-author network
python phase2.py --orcid 0000-0000-0000-0000 --trajectory

# Phase 3: Full analysis (recommended — includes Phases 1 + 2)
python phase3.py --orcid 0000-0000-0000-0000 --trajectory

# Also accepts OpenAlex IDs
python phase3.py --openalex-id A0000000000 --trajectory
\end{verbatim}

\subsubsection{Common Flags (Table \ref{tab:common-flags})}\label{sec:hercon-weight-customization-2}

\begin{table}[H]
\centering
\begin{tabular}{ll}
\toprule
\textbf{Flag} & \textbf{Description} \\
\midrule
\texttt{--export results.json} & Export summary to JSON \\
\texttt{--trajectory} or \texttt{-t} & Show cumulative career trajectory by year \\
\texttt{--since 2010} & Only include works from 2010 onward \\
\texttt{--depth 1|2|3} & Co-author graph depth (default 2) \\
\texttt{--herocon-weights file.json} & Custom HEROCON weight configuration \\
\texttt{--confirm} or \texttt{-c} & Review ORCID-flagged works interactively before scoring \\
\texttt{--no-orcid-check} & Skip ORCID cross-validation \\
\texttt{--no-audit} & Skip audit file generation (not recommended) \\
\texttt{--verbose} or \texttt{-v} & Verbose output \\
\bottomrule
\end{tabular}
\caption{Common flags for command-line interface}
\label{tab:common-flags}
\end{table}

\subsubsection{Interactive Confirmation Mode}

\begin{verbatim}
python phase3.py --orcid 0000-0000-0000-0000 --confirm
\end{verbatim}

The tool displays flagged works with reasons and prompts for a decision. Input options: \texttt{all} (exclude all), \texttt{none} (keep all), \texttt{1,3,5} (exclude specific items), or \texttt{1-3,5} (ranges). The audit trail records all exclusion decisions.

\subsubsection{Audit Trail}

Every run generates a timestamped JSON audit file in \texttt{./audits/} by default. CLI-generated audit files can be uploaded to the web interface for interactive visualization without re-running computation. Note: the \texttt{--trajectory} or \texttt{-t} flag must be included during generation for the career trajectory chart to be available in the web interface. All other visualizations work regardless of this flag.

\subsection{Running Locally}

\textbf{Option A: Python.} 

\begin{verbatim}
cd citation-constellation/
pip install -r requirements.txt
python app/main.py
# Open http://localhost:7860
\end{verbatim}

\textbf{Option B: Docker from source.}
\begin{verbatim}
docker build --platform linux/amd64 -t citation-constellation:v0.3 .
docker run --rm -it -p 7860:7860 citation-constellation:v0.3
# Open http://localhost:7860
\end{verbatim}

\textbf{Option C: Prebuilt image.}
\begin{verbatim}
docker pull mahbub1969/citation-constellation:v1
docker run --rm -it -p 7860:7860 mahbub1969/citation-constellation:v1
# Open http://localhost:7860
\end{verbatim}

\subsection{Citation-Cosmograph Ecosystem (Table \ref{tab:citation-cosmograph-ecosystem})}

\begin{center}
\texttt{pulsar} $\rightarrow$ \texttt{astrolabe} $\rightarrow$ \texttt{constellation}\\[2pt]
{\small\textit{the signal \hspace{1.2em} the instrument \hspace{1.2em} the map}}
\end{center}

\begin{table}[H]
\centering
\small
\begin{tabular}{@{}lll@{}}
\toprule
\textbf{Component} & \textbf{Purpose} & \textbf{Link} \\
\midrule
Citation-Constellation
  & BARON \& HEROCON scoring
  & \href{https://citation-constellation.serve.scilifelab.se}{\textbf{No-Code Tool}} {\footnotesize|} \href{https://github.com/citation-cosmograph/citation-constellation}{Code} \\
Citation-Pulsar-Helm
  & LLM inference on Kubernetes
  & \href{https://github.com/citation-cosmograph/citation-pulsar-helm}{Code} \\
Citation-Astrolabe
  & Venue governance database
  & \href{https://github.com/citation-cosmograph/citation-astrolabe}{Code} \\
\bottomrule
\end{tabular}
\caption{\href{https://github.com/citation-cosmograph}{The Citation-Cosmograph ecosystem on GitHub.}}
\label{tab:citation-cosmograph-ecosystem}
\end{table}


\section{Demonstration and Illustrative Output}

\subsection{Score Progression Across Phases}

As detection layers deepen (Table \ref{tab:score-progression}), BARON decreases as more in-group relationships are identified:

\begin{table}[H]
\centering
\begin{tabular}{lllll}
\toprule
\textbf{Phase} & \textbf{Detection Layers} & \textbf{Classes} & \textbf{BARON} & \textbf{HEROCON} \\
\midrule
1 & Self-citation & 2 & High & --- \\
2 & + Co-author network & 4 & Moderate & Moderate-High \\
3 & + Affiliation matching & 7 & Lower & Moderate \\
\bottomrule
\end{tabular}
\caption{Score progression across phases}
\label{tab:score-progression}
\end{table}

\paragraph{Ethical Notice}

\begin{figure}[H]
\centering
\includegraphics[width=\textwidth]{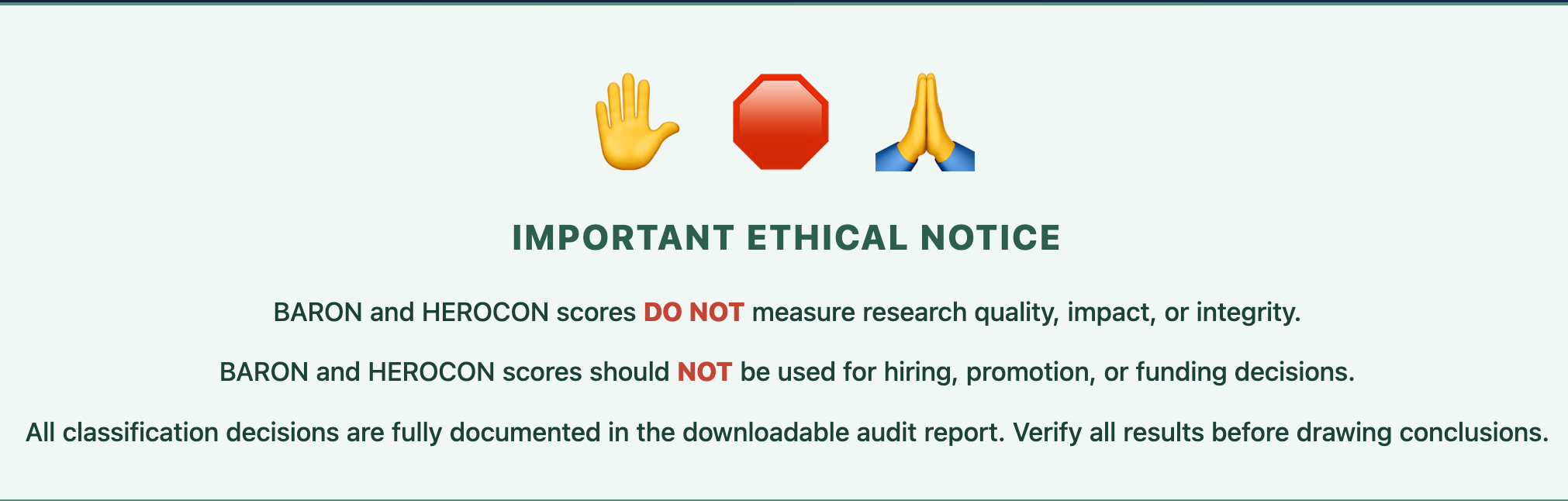}
\caption{Ethical notice displayed at the top of every analysis output.}
\label{fig:notice}
\end{figure}

Every analysis output begins with a prominent ethical disclaimer (Figure \ref{fig:notice}), reinforcing that BARON and HEROCON measure citation network structure, not research quality, impact, or integrity, and should not be used for hiring, promotion, or funding decisions.

\paragraph{Full Score Panel}

\begin{figure}[H]
\centering
\begin{subfigure}[b]{\textwidth}
    \centering
    \includegraphics[width=\textwidth]{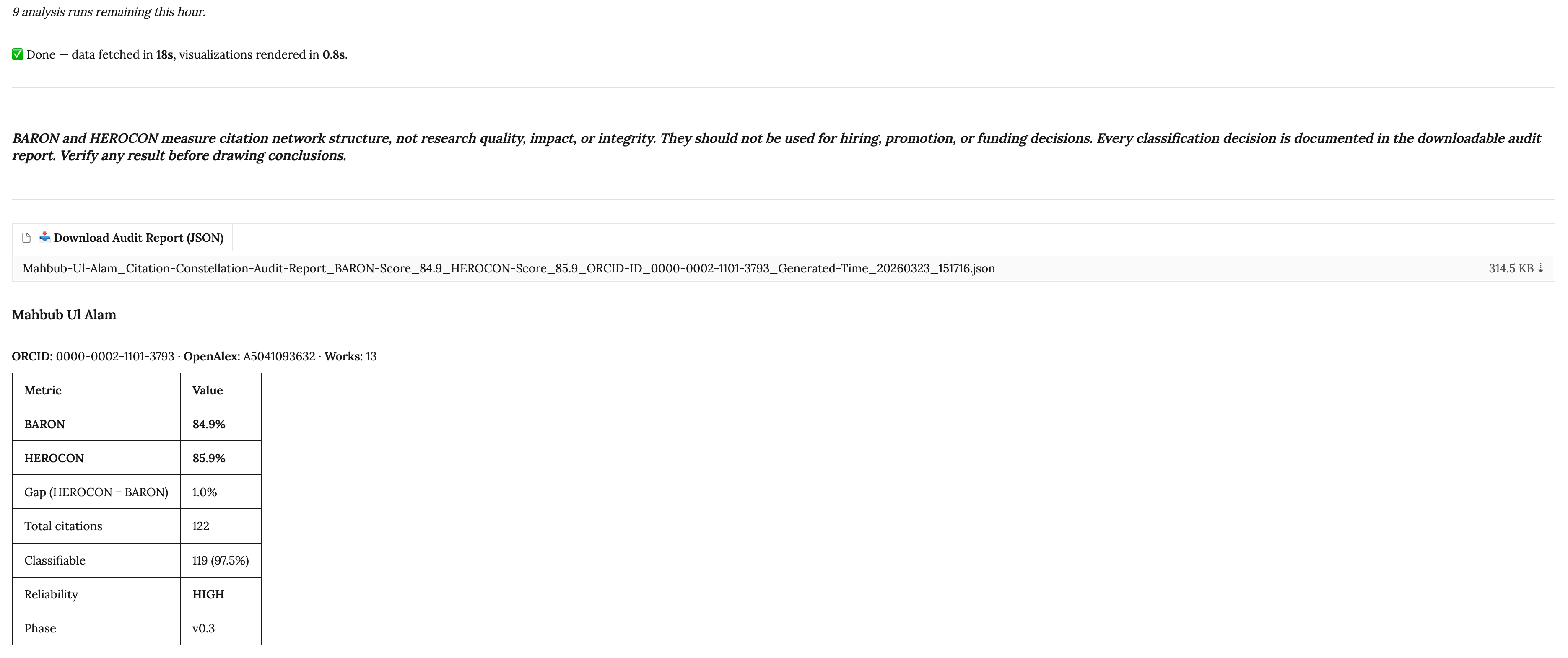}
    \caption{Web Interface}
\end{subfigure}

\begin{subfigure}[b]{\textwidth}
    \centering
    \includegraphics[width=\textwidth]{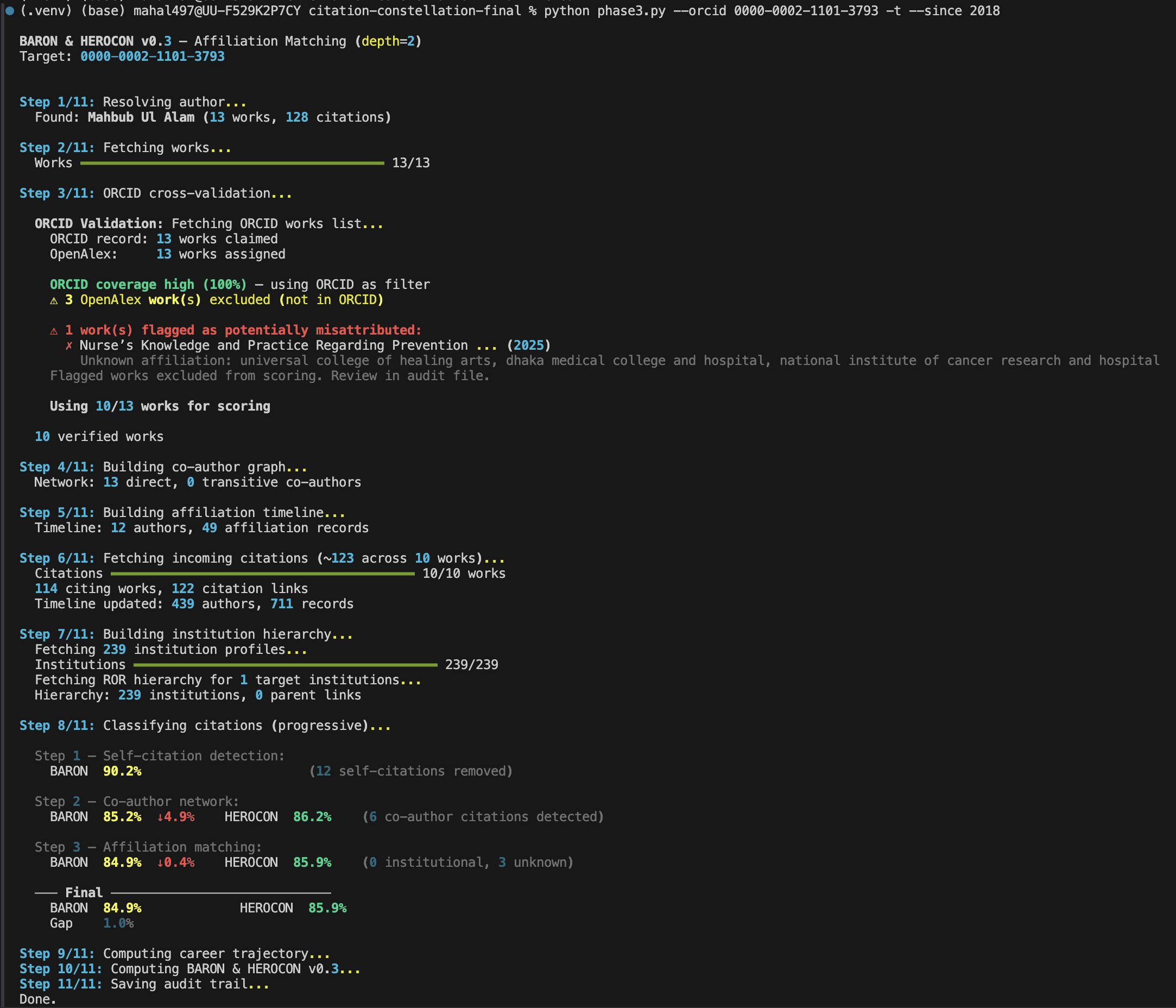}
    \caption{Command Line Interface}
\end{subfigure}
\caption{Score panel in both interfaces.}
\label{fig:score-panel}
\end{figure}

The score panel (Figure \ref{fig:score-panel}) presents the BARON and HEROCON scores alongside key summary statistics: total citations, classifiable citations, the BARON--HEROCON gap, and a data quality reliability rating. Both the web interface and command-line interface produce equivalent information, formatted for their respective contexts.

\paragraph{Classification Breakdown Donut Chart}

\begin{figure}[H]
\centering
\includegraphics[width=0.5\textwidth]{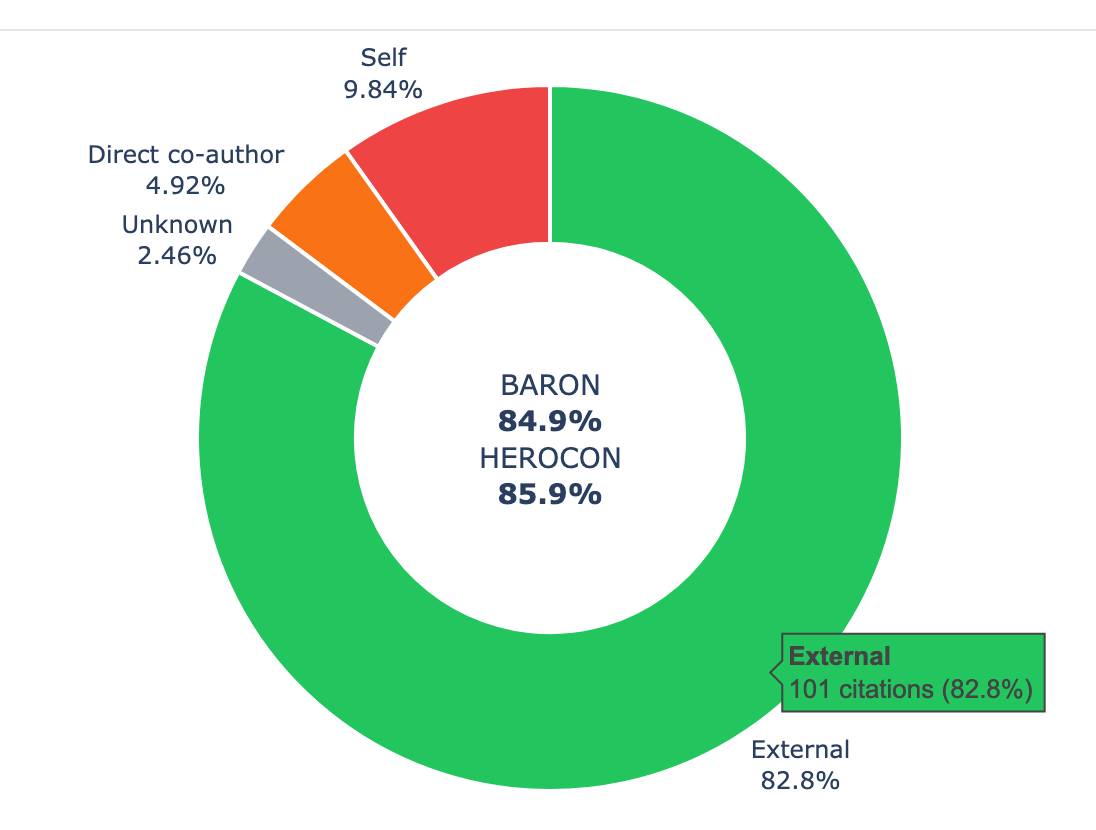}
\caption{Classification breakdown donut chart.}
\label{fig:donut-chart}
\end{figure}

The donut chart (Figure \ref{fig:donut-chart}) provides a proportional breakdown of citation origins across all classification categories, with the BARON and HEROCON scores displayed in the center. This gives an immediate visual sense of how a researcher's citations distribute across network layers.

\paragraph{Co-author Network Graph}

\begin{figure}[H]
\centering
\includegraphics[width=\textwidth]{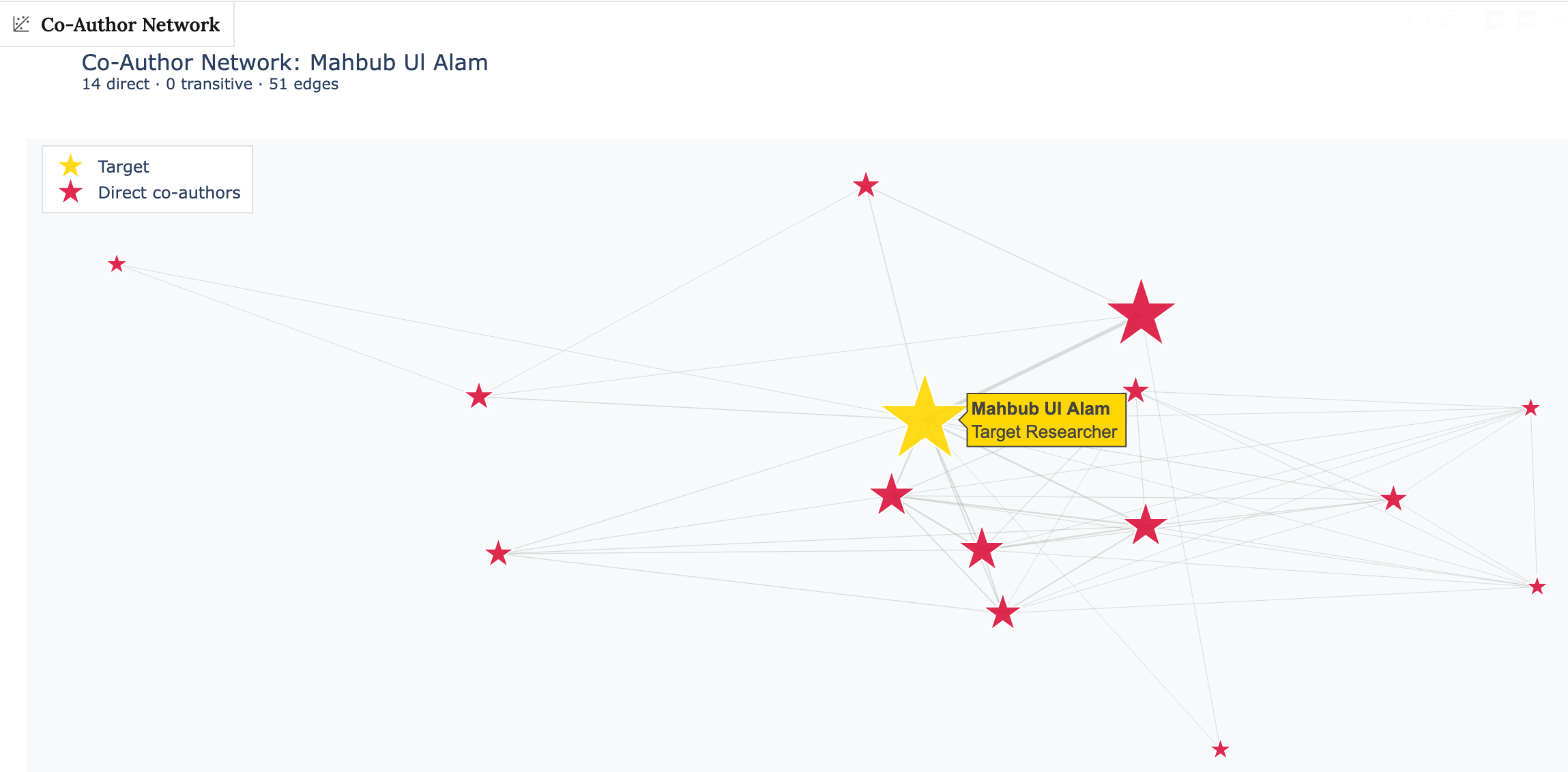}
\caption{Co-author network graph (overview).}
\label{fig:co-author-network-graph-1}
\end{figure}

\begin{figure}[H]
\centering
\includegraphics[width=\textwidth]{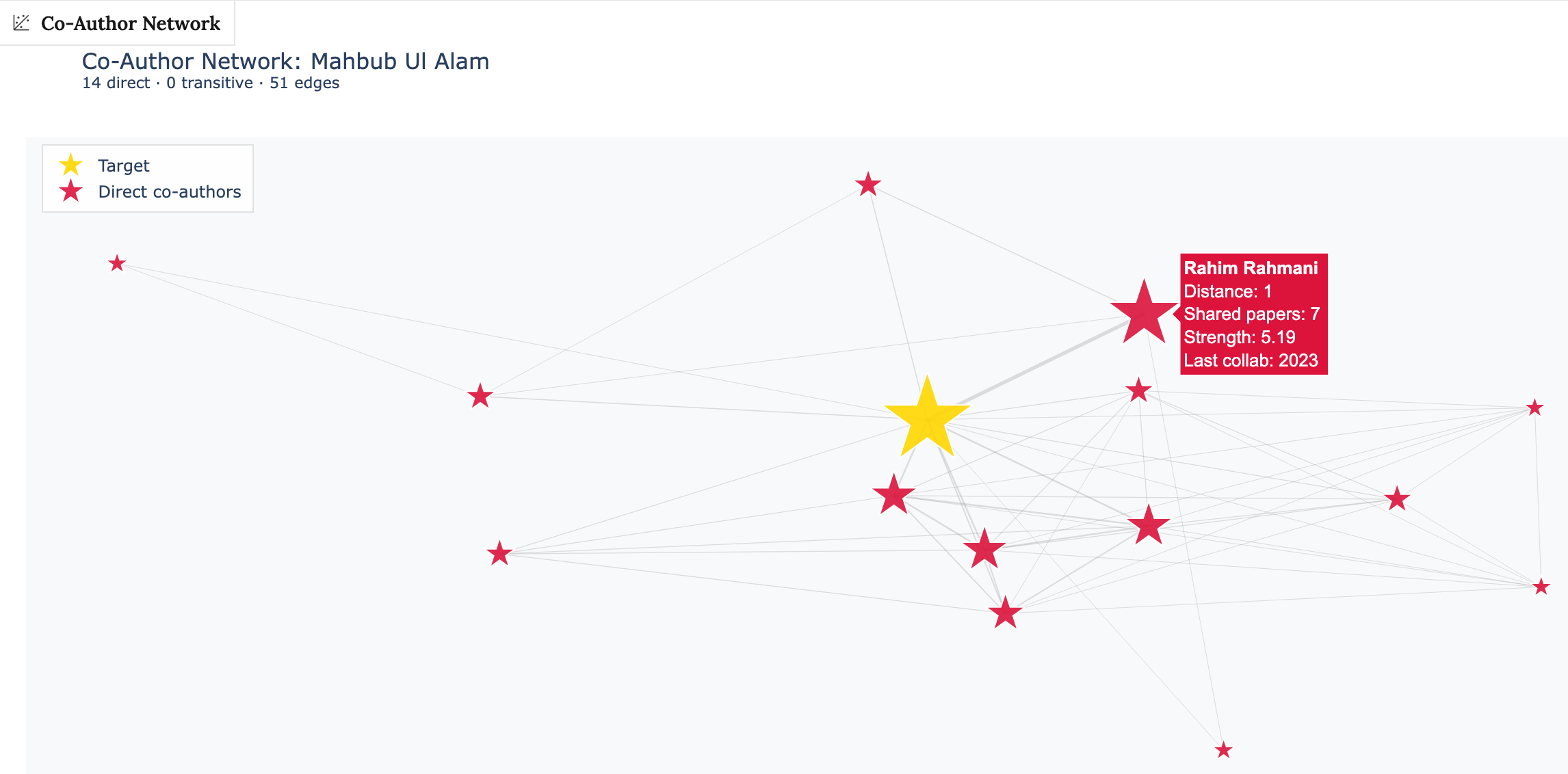}
\caption{Co-author network graph (detail).}
\label{fig:co-author-network-graph-2}
\end{figure}

The interactive co-author network graph (Figures \ref{fig:co-author-network-graph-1} and \ref{fig:co-author-network-graph-2}) renders the target researcher as a gold node, direct co-authors in crimson (sized proportionally to the number of shared publications), and transitive co-authors in blue. Hovering over any node reveals the author name, shared paper count, and co-authorship recency. Networks exceeding 150 nodes are automatically pruned for readability.

\paragraph{Classification Summary}

\begin{figure}[H]
\centering
\begin{subfigure}[b]{\textwidth}
    \centering
    \includegraphics[width=\textwidth]{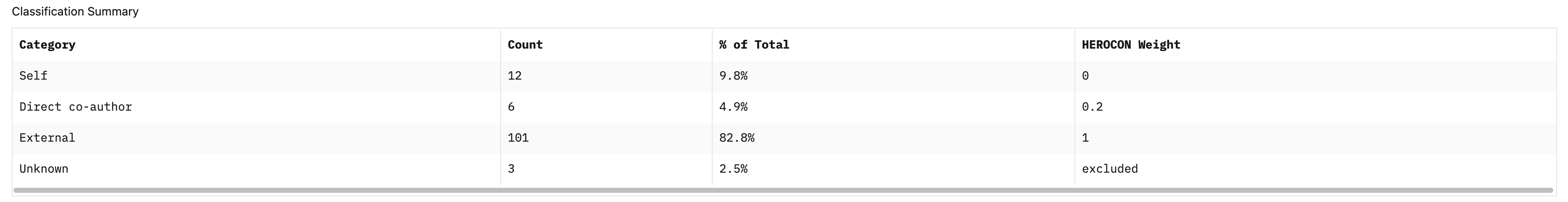}
    \caption{Web Interface}
\end{subfigure}

\begin{subfigure}[b]{\textwidth}
    \centering
    \includegraphics[width=\textwidth]{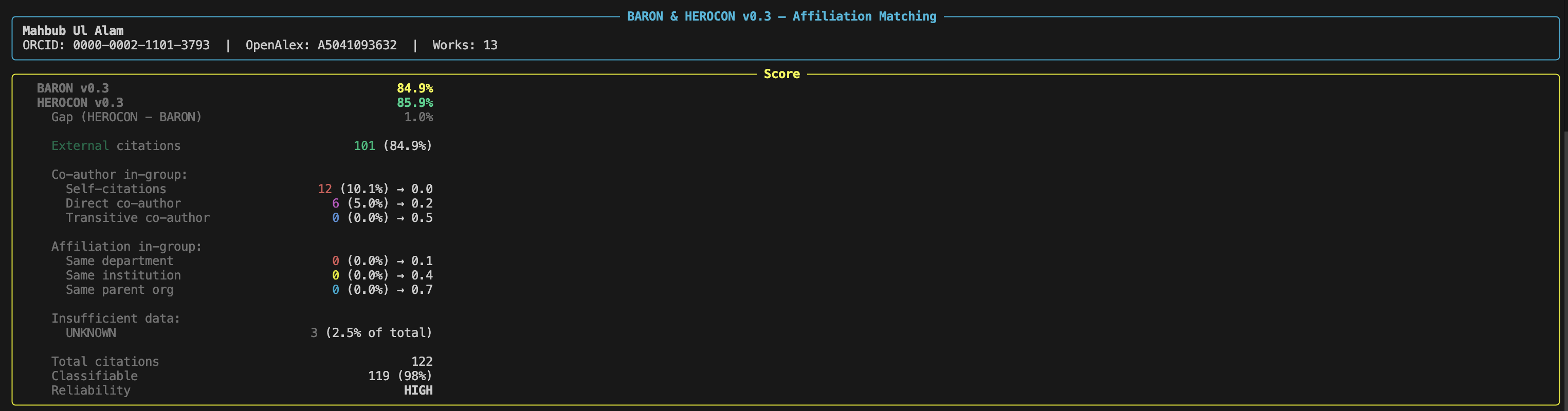}
    \caption{Command Line Interface}
\end{subfigure}
\caption{Classification summary in both interfaces.}
\label{fig:classification-summary}
\end{figure}

The classification summary table (Figure \ref{fig:classification-summary}) lists each citation category with its count, percentage of classifiable citations, and the HEROCON weight applied. This tabular view complements the donut chart by providing exact numbers for each network layer.

\paragraph{Career Trajectory Chart}

\begin{figure}[H]
\centering
\begin{subfigure}[b]{\textwidth}
    \centering
    \includegraphics[width=\textwidth]{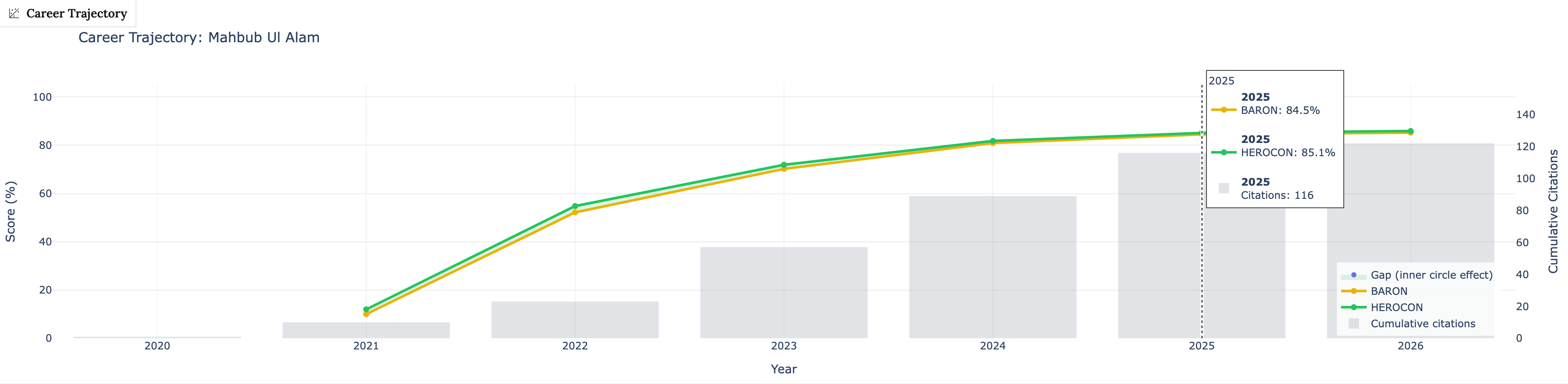}
    \caption{Web Interface}
\end{subfigure}

\begin{subfigure}[b]{\textwidth}
    \centering
    \includegraphics[width=\textwidth]{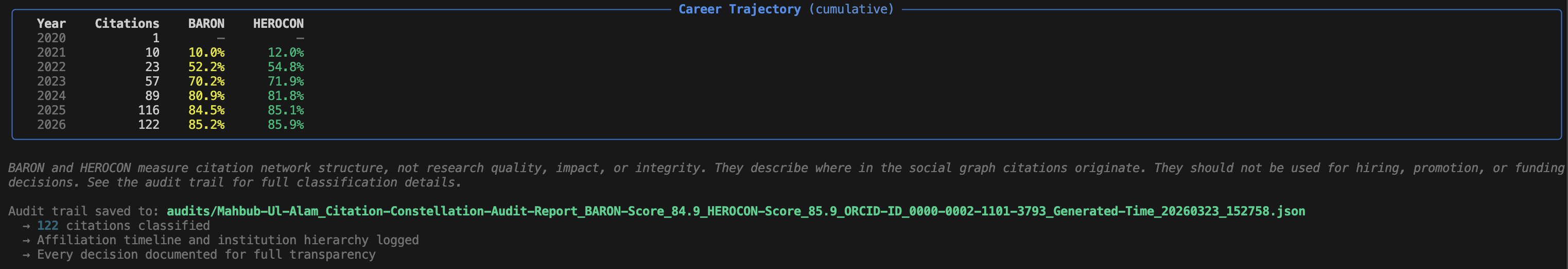}
    \caption{Command Line Interface}
\end{subfigure}
\caption{Career trajectory chart in both interfaces.}
\label{fig:career-trajectory-chart}
\end{figure}

The career trajectory chart (Figure \ref{fig:career-trajectory-chart}) plots cumulative BARON and HEROCON scores over time as dual lines, with a shaded region between them representing the gap. Stacked bars beneath show annual citation volume. This visualization reveals how a researcher's external reach evolves across career stages, whether it grows after an institutional move, narrows as a lab expands, or remains stable over time.

\paragraph{Citation Table from Audit Trail}

\begin{figure}[H]
\centering
\includegraphics[width=\textwidth]{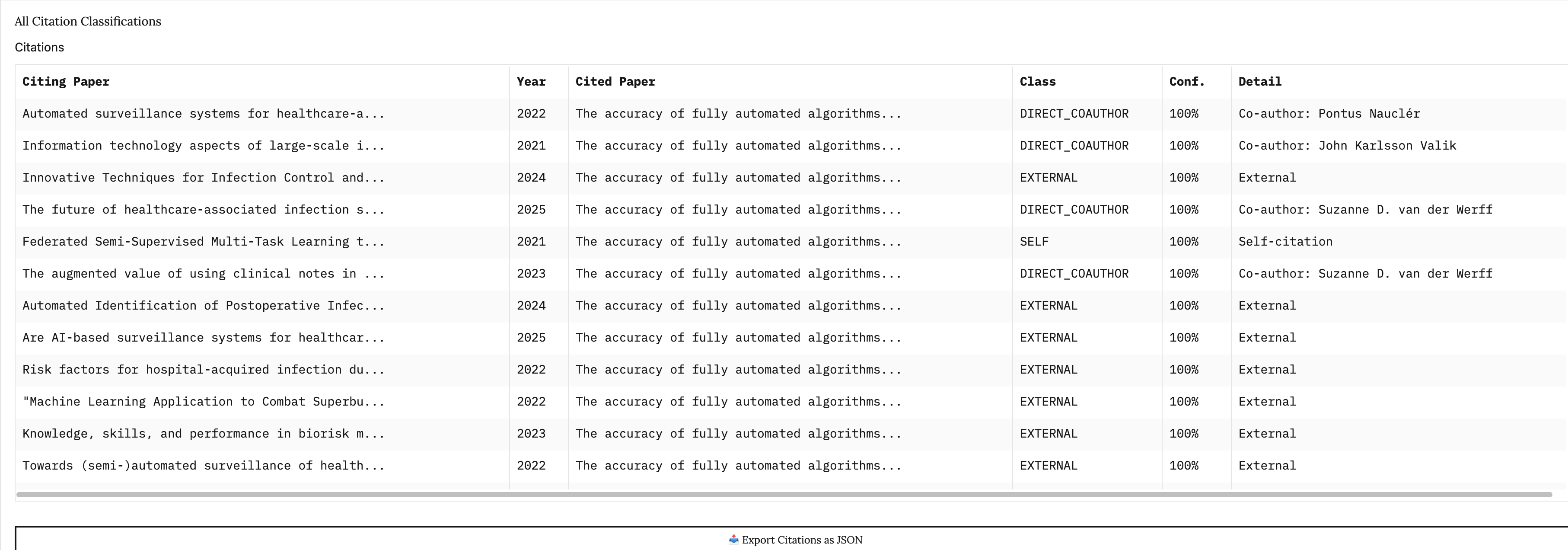}
\caption{Full citation table from the audit trail.}
\label{fig:citation-table}
\end{figure}

The full citation table (Figure \ref{fig:citation-table}) exposes every individual citation with its classification, confidence level, detection phase, and a human-readable rationale explaining why that classification was assigned. This is the audit trail made visible: any classification can be inspected, questioned, and contested.

\paragraph{Comparison View}

\begin{figure}[H]
\centering
\includegraphics[width=\textwidth]{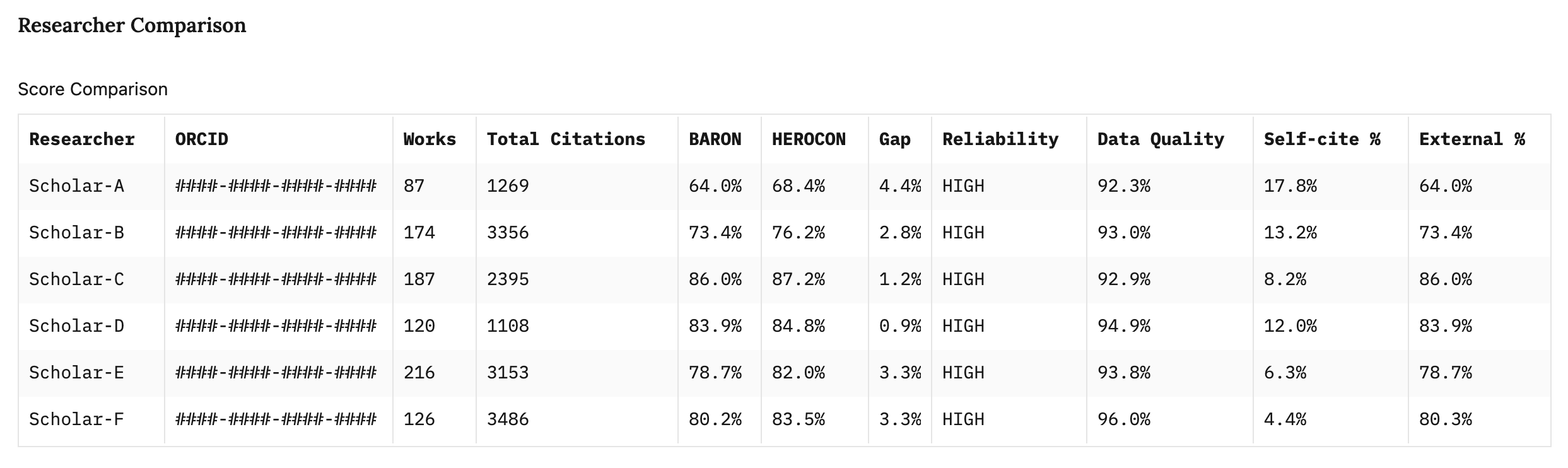}
\caption{Comparison table.}
\label{comparison-table}
\end{figure}

\begin{figure}[H]
\centering
\includegraphics[width=\textwidth]{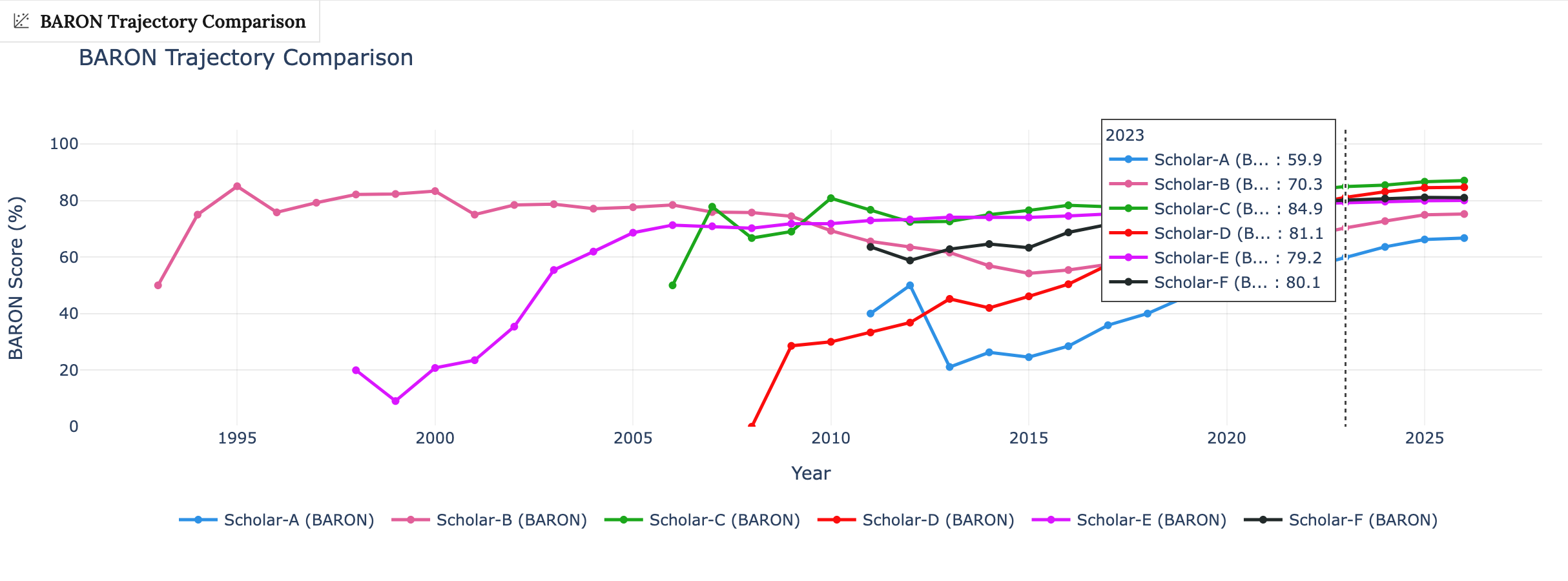}
\caption{BARON trajectory comparison.}
\label{BARON-trajectory-comparison.}
\end{figure}

\begin{figure}[H]
\centering
\includegraphics[width=\textwidth]{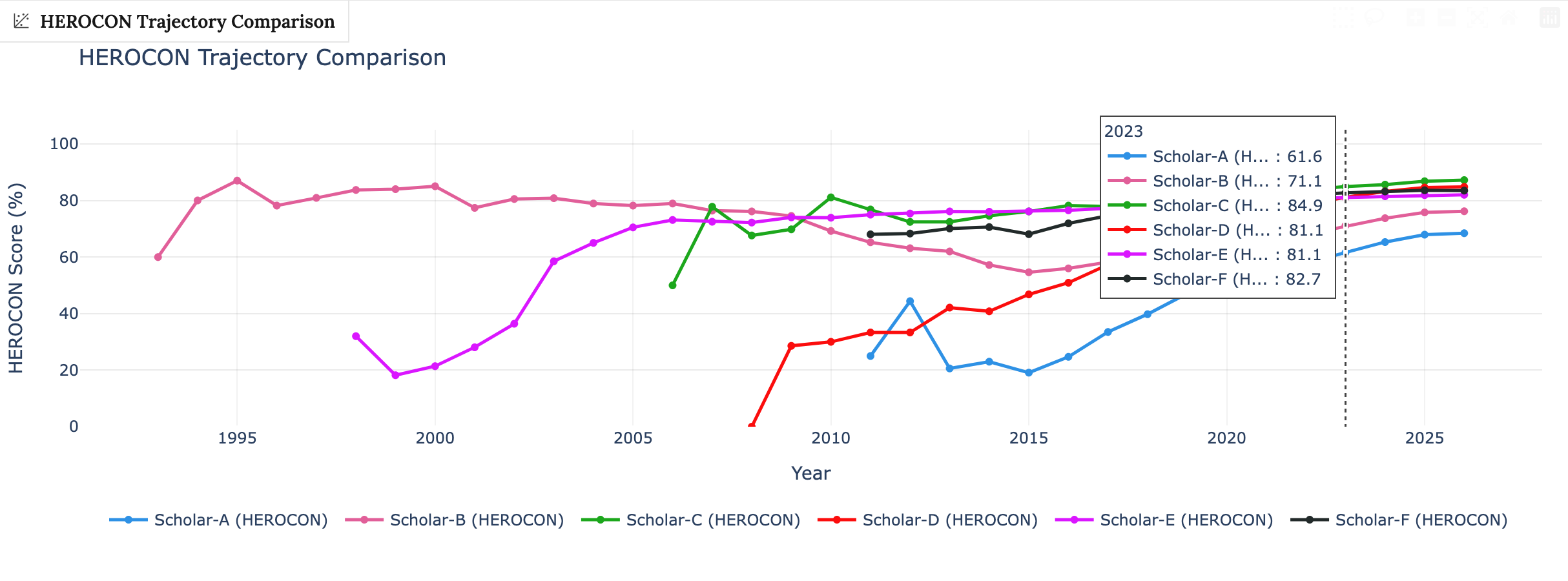}
\caption{HEROCON trajectory comparison.}
\label{HEROCON-trajectory-comparison.}
\end{figure}

The comparison view enables side-by-side structural analysis of multiple researchers from uploaded audit files. The tables and figures (Figures \ref{comparison-table}, \ref{BARON-trajectory-comparison.}, and \ref{HEROCON-trajectory-comparison.}) presented here are derived from actual empirical data; however, researcher names have been anonymized to ensure confidentiality.

The overlaid trajectory charts allow structural comparison of how external reach evolves across different researchers' careers. A department head could use these to understand whether early-career researchers in their group are building external visibility at a pace consistent with field norms, or whether a mid-career shift in collaboration patterns coincided with a change in citation composition. When researchers from different fields or institution types are compared, the trajectories can reveal whether apparent differences in raw citation counts mask similar underlying structural patterns, or whether nominally similar profiles diverge sharply in their network dependence over time. These are not evaluative rankings; they are structural narratives placed side by side.

\begin{figure}[H]
\centering
\includegraphics[width=\textwidth]{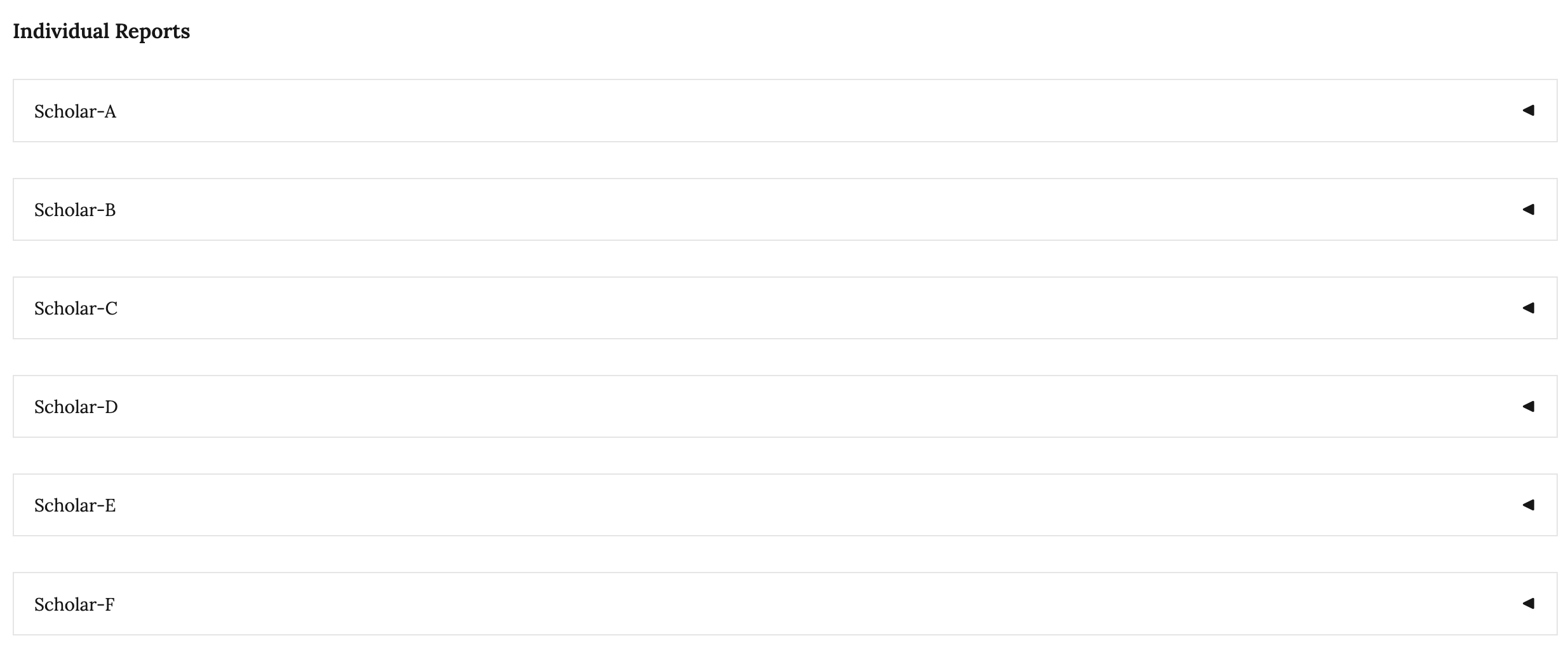}
\caption{Individual reports within the comparison view.}
\label{visualize-individual-reports}
\end{figure}

In addition to the comparative overlays, the visualization expands into full individual reports (Figure \ref{visualize-individual-reports}) for each uploaded researcher, identical in format to those produced by the Run Analysis tab. Each expandable section contains the score panel, classification donut chart, co-author network graph, career trajectory, and full citation table for that researcher. This allows a viewer to move seamlessly between the bird's-eye comparative view and the detailed per-citation audit of any individual profile, without needing to re-run any computation.

\subsection{ORCID Cross-Validation Impact}

In testing, OpenAlex had merged works from a different researcher with a similar name. The ORCID layer correctly excluded those misattributed works, preventing contamination of the co-author graph and affiliation timeline. DOI-based matching resolved 70--85\% of works; title-based fuzzy matching recovered an additional 10--15\%.

\subsection{Limitations of Current Validation}

I acknowledge that the current demonstration is descriptive rather than validational. I have not yet established that BARON/HEROCON scores correlate with (or meaningfully diverge from) independent measures of citation motivation, research quality, or integrity. I identify such validation as a critical priority for future work (see Section~\ref{sec:future-work}).

\section{Nomenclature as Framework: The Metaphors of BARON and HEROCON}\label{sec:metaphor_meaning}

\begin{figure}[h]
   \centering
    \includegraphics[width=\textwidth]{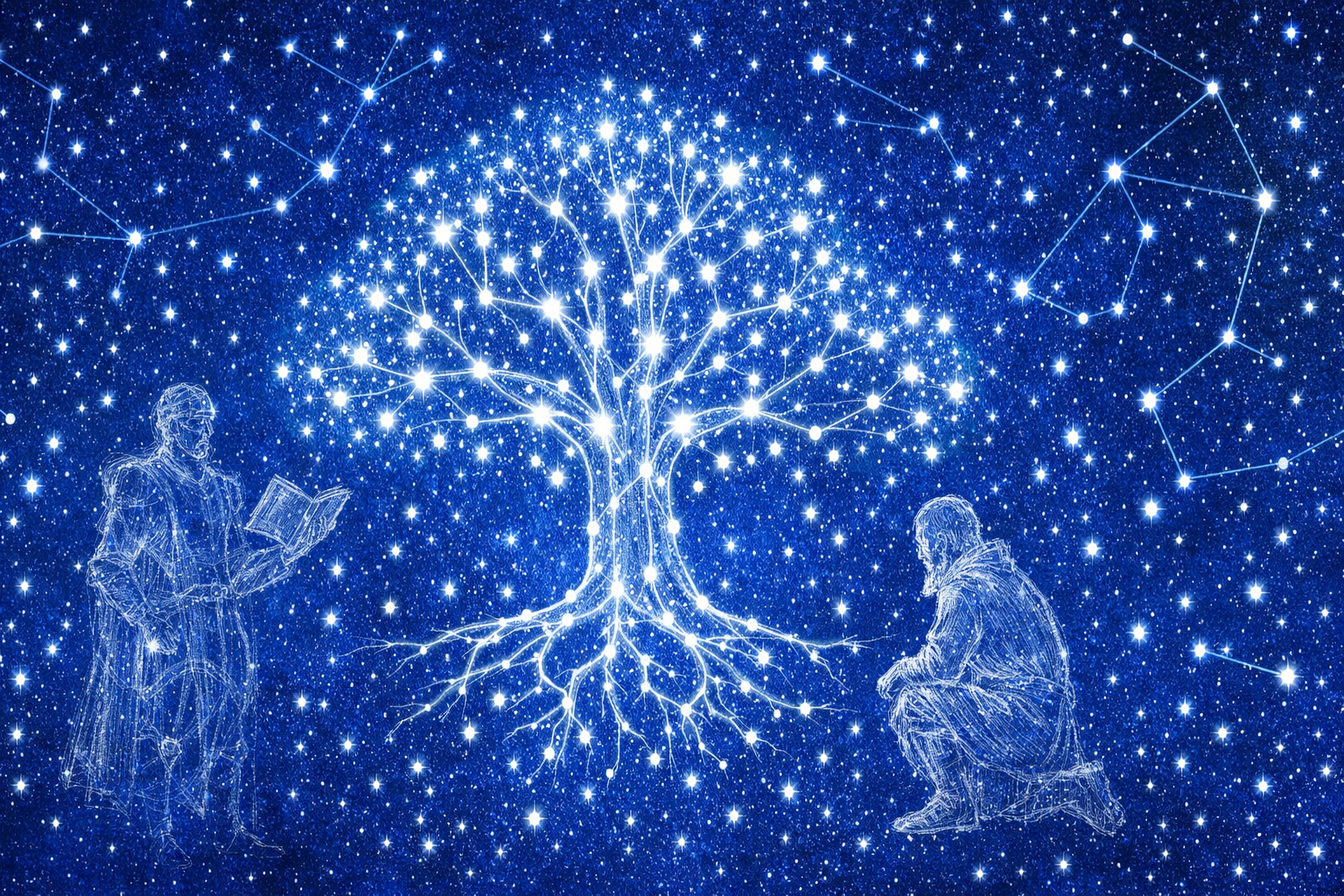}
    \caption{The Citation-Constellation emblem}
    \label{fig:logo}
\end{figure}

Before describing the technical methodology, I pause to explain the names of the two scores, because the metaphors they invoke are not decorative. They encode the conceptual relationship (Figure \ref{fig:logo}) between the scores and frame how they should be interpreted.

\subsection{BARON: The Boundary Guardian}

The Boundary-Anchored Research Outreach Network (BARON) score takes its name from the historical \textit{Marcher Barons}, the feudal lords charged with securing and governing the outer borders of a realm. In medieval England and Wales, the Marcher Lords held the frontier: the boundary where a kingdom's influence either held or dissolved. Their role was to anchor legitimacy at the edges.

The BARON score plays an analogous role for a researcher's citation profile. It deliberately filters out the natural amplification of local networks, co-authors, institutional peers, and editorial connections, and measures only what crosses the boundary: citations from researchers with no detected relationship to the author. In doing so, BARON anchors the maximum defensible reach of a scholar's metrics. It asks a simple, strict question: \textit{how much of this researcher's citation impact survives when we remove every citation that could plausibly have been mediated by social proximity?}

This is a conservative measure by design. Just as a Marcher Baron's strength was tested not by the loyalty of subjects within the walls but by the respect commanded beyond them, BARON tests a citation profile not by in-group endorsement but by external validation. A high BARON score does not mean a researcher is better; it means their work has demonstrably reached beyond their immediate scholarly community. A low BARON score does not mean a researcher is worse; it may reflect a small field, a productive lab, or a disciplinary norm of close collaboration.

By establishing this foundational threshold of strict external outreach, the BARON score provides the crucial boundary that gives the broader HEROCON score its shape. Without demonstrated reach beyond the immediate circle, the constellation cannot form.

\subsection{HEROCON: The Constellation}

The Holistic Equilibrated Research Outreach CONstellation (HEROCON) score is named for the constellation Hercules. In Greek mythology, Hercules was placed among the stars after his death as an eternal monument to strength, perseverance, and labors that transcended mortal limits.

Where BARON draws a binary line at the boundary, HEROCON maps the full figure. It treats a researcher's total scholarly influence as a constellation: localized collaborations, co-authors, departmental colleagues, and institutional peers form bright, dense clusters of stars, much like the tightly bound star systems at the heart of Hercules. These clusters are valuable; they represent the close intellectual community that sustains and nourishes research. But a constellation cannot be confined to a single cluster. It must stretch across the sky to earn its name.

The BARON score provides the anchoring boundary stars that define the constellation's outer limits and give it its legendary shape. A dense local cluster without distant anchor points is not a constellation; it is an asterism, a local pattern that does not extend to the broader sky. Conversely, scattered distant points without a bright local core lack the structure that makes a constellation recognizable. The interplay between the two, dense local community (HEROCON's partial credit for in-group citations) and broad external reach (BARON's boundary anchoring), is what gives a citation profile its constellation shape.

One of the constellation's most prominent star systems, \textit{Rasalgethi} (Alpha Herculis, from the Arabic \textit{ra's al-jāthī}, "the kneeler's head"), carries a further resonance.

The figure of Hercules in the sky is depicted kneeling, a posture not of defeat but of humility. True scholarly leadership, like the kneeling Hercules, requires humility: acknowledging the contributions of one's immediate community, recognizing that co-author citations and institutional support are the bright core that sustains a career. A high HEROCON score demonstrates precisely this, that a researcher unites a bright local foundation with the vast external outreach that illuminates the broader academic universe.

Ultimately, the gap between HEROCON and BARON reveals the shape of the constellation itself. A small gap means the constellation is diffuse, citations come from everywhere. A large gap means the constellation has a bright, dense core surrounded by more distant anchor points. Neither pattern is inherently better; they are different shapes in the scholarly sky, and the Citation-Constellation tool makes those shapes visible.

\section{Methodology}

\subsection{Conceptual Framework}

I conceptualize a researcher's citation profile as concentric network layers (Table \ref{tab:conceptualizing-scoring}), from most proximate (self) to most distant (external). Each layer represents a relationship type that could mediate citation behavior:

\begin{itemize}
    \item \textbf{Layer 0 --- Self:} Citing own prior work.
    \item \textbf{Layer 1 --- Direct co-authors:} Shared at least one publication.
    \item \textbf{Layer 2 --- Transitive co-authors:} Co-authors of co-authors (configurable depth, default 2).
    \item \textbf{Layer 3 --- Institutional colleagues:} Same institution/department, even without co-authorship.
    \item \textbf{Layer 4 --- Venue governance:} Editorial board or program committee overlap with citing venue.
    \item \textbf{Layer 5 --- External:} No detected relationship.
\end{itemize}

BARON treats all layers 0--4 as in-group (weight = 0) and only counts fully external citations (weight = 1). HEROCON assigns graduated weights:

\begin{table}[H]
\centering
\small
\begin{tabular}{lllll}
\toprule
\textbf{Layer} & \textbf{Relationship} & \textbf{BARON} & \textbf{HEROCON} & \textbf{Rationale} \\
\midrule
0   & Self-citation                     & 0 & 0.0  & No credit; researcher is citing themselves \\
1   & Direct co-author                  & 0 & 0.2  & Low credit; strongest collaborative tie \\
2   & Transitive co-author              & 0 & 0.5  & Moderate credit; indirect tie, weaker influence pathway \\
3a  & Same department                   & 0 & 0.1  & Very low credit; strongest proximity without co-authorship \\
3b  & Same institution, different dept   & 0 & 0.4  & Moderate credit; cross-departmental \\
3c  & Same parent organization          & 0 & 0.7  & High credit; tenuous institutional link \\
4a  & Venue self-governance             & 0 & 0.05 & Near-zero; researcher directly governs the venue \\
4b  & Venue editor is co-author         & 0 & 0.15 & Low credit; compound relationship \\
4c  & Venue editor at same institution  & 0 & 0.3  & Moderate credit; institutional link through editorial channel \\
4d  & Committee member in network       & 0 & 0.4  & Moderate credit; weaker governance connection \\
5   & External                          & 1 & 1.0  & Full credit; no detected relationship \\
\bottomrule
\end{tabular}
\caption{Conceptualizing scoring for a researcher's citation profile as concentric network layers.}
\label{tab:conceptualizing-scoring}
\end{table}

\paragraph{Score computation.}

\begin{equation}
\text{BARON} = \frac{\text{external citations}}{\text{classifiable citations}} \times 100
\end{equation}

\begin{equation}
\text{HEROCON} = \frac{\sum_{i} w_i}{\text{classifiable citations}} \times 100
\end{equation}

\noindent where $w_i$ is the HEROCON weight for citation $i$'s classification, and ``classifiable'' excludes UNKNOWN citations (see Section~\ref{sec:unknown}).

\begin{equation}
\text{Diagnostic gap} = \text{HEROCON} - \text{BARON}
\end{equation}

\noindent The diagnostic gap represents the proportion of impact attributable to in-group citations under graduated weighting.

\subsubsection{The HEROCON weights as a testable hypothesis}\label{sec:hercon-weight-discussion}

The HEROCON weights represent a formal hypothesis about the relative strength of different network pathways in mediating citation. I hypothesize that a direct co-authorship tie (weight 0.2) is a stronger predictor of network-mediated citation than mere departmental colocation (weight 0.1), because the former represents a deliberate intellectual collaboration while the latter may be purely administrative. I hypothesize that a tie through a co-author of a co-author (weight 0.5) is weaker than a direct tie but still represents a meaningful intellectual community, whereas a different-department colleague at the same institution (weight 0.4) might have even less direct intellectual overlap despite closer physical proximity. These hypotheses are eminently testable, and I identify their empirical calibration as my primary future work (see Section~\ref{sec:future-work}).

I acknowledge these weights are experimental defaults, not empirically calibrated values. Researchers who wish to test alternative weightings may do so through full weight customization, available both via the CLI (\texttt{--herocon-weights path/to/weights.json}, see Section~\ref{sec:hercon-weight-customization-2} for details) and through the web interface's Advanced settings, where a custom weights JSON file can be uploaded before running an analysis (see Section~\ref{sec:hercon-weight-customization-1}, step \textbf{5} for details).

\begin{figure}[H]
   \centering
   \includegraphics[scale=.2]{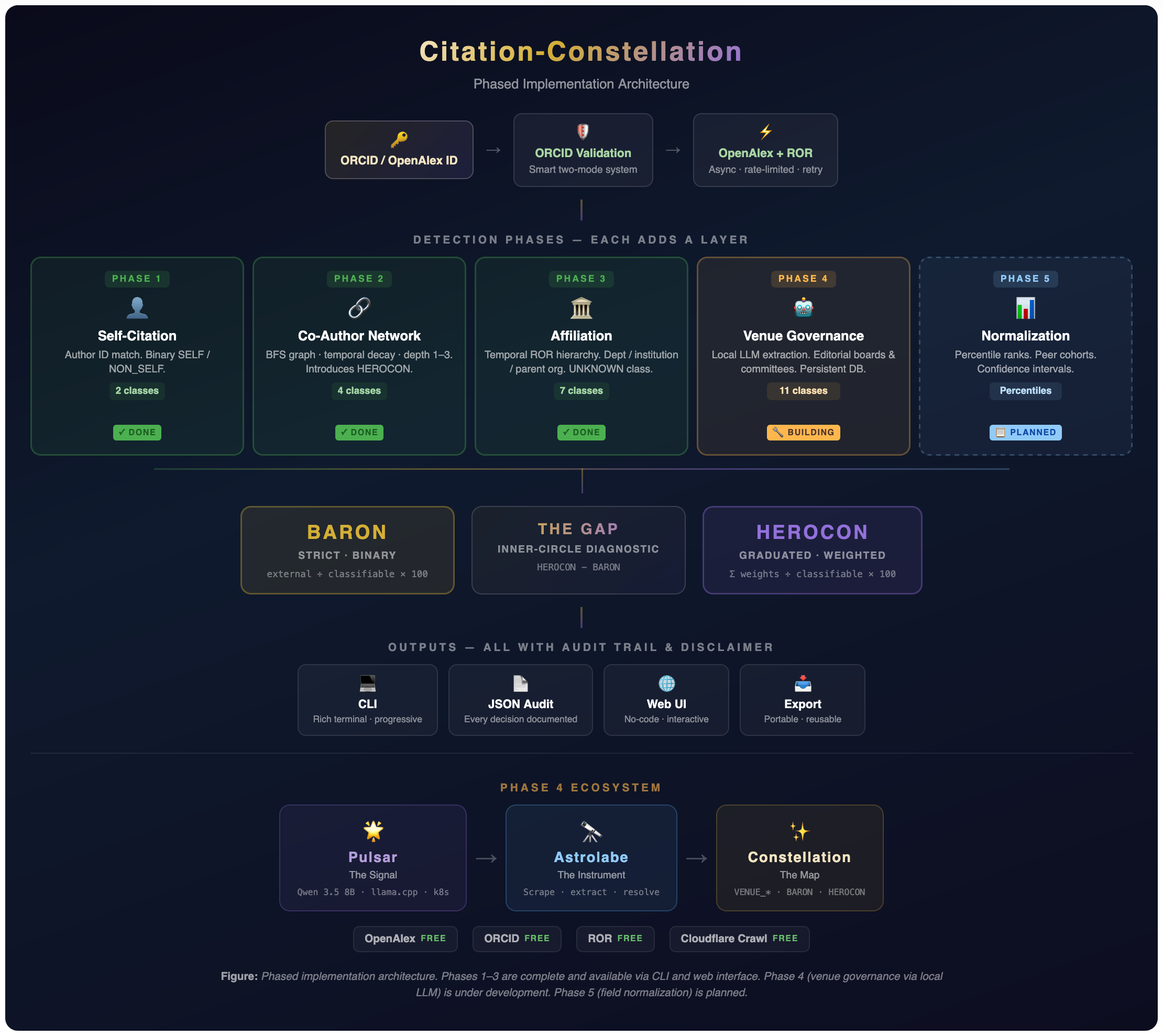}
    \caption{Phased implementation architecture. Phases 1–3 are complete and available via CLI and web interface. Phase 4 (venue governance via local LLM) is under development. Phase 5 (field normalization) is planned.}
    \label{contribution}
\end{figure}

\subsection{Phased Implementation Architecture (Figure \ref{contribution})}

Each phase adds a detection layer and produces a usable score. Later phases incorporate earlier layers.

\subsubsection{Phase 1: Self-Citation Baseline}

For each researcher publication, I fetch all incoming citations from OpenAlex. I check whether any author ID on the citing work matches the target researcher and classify as \texttt{SELF} or \texttt{NON\_SELF}.

\begin{equation}
\text{BARON}_{v0.1} = \text{percentage of \texttt{NON\_SELF} citations}
\end{equation}

\subsubsection{Phase 2: Co-Author Network Detection}

I construct a co-authorship graph from the target researcher's publications. Each author pair sharing a paper creates a weighted edge:

\begin{equation}
\text{strength}(a, b) = \text{shared\_papers}(a, b) \times \exp(-0.1 \times \text{years\_since\_last\_collaboration})
\end{equation}

The exponential decay with rate 0.1 yields a half-life of approximately 7 years ($\ln(2)/0.1 \approx 6.93$). This reflects the hypothesis that recent collaborators are more ``in-group'' than distant ones. I note that the specific decay rate is not empirically grounded; alternative rates would be equally defensible. The decay function currently modulates co-authorship strength metadata logged in audit files but does not affect BARON classification (which uses binary in/out).

Starting from the target researcher, the tool performs a Breadth-First Search (BFS) traversal of the co-authorship graph to a configurable depth $d$ (default 2). Every author encountered during this traversal is considered part of the researcher's collaborative network. Citations are then classified by graph distance: distance 0 (the researcher themselves) maps to \texttt{SELF}, distance 1 (direct co-authors) maps to \texttt{DIRECT\_COAUTHOR}, distance 2 (co-authors of co-authors) maps to \texttt{TRANSITIVE\_COAUTHOR}, and any citing author beyond depth $d$ or not found in the graph is classified as \texttt{EXTERNAL}.

HEROCON is introduced in this phase with the 4-class taxonomy.

\subsubsection{Phase 3: Temporal Affiliation Matching}

Citations not classified as \texttt{SELF} or co-author are checked for institutional affiliation overlap at the time of citation. I build an affiliation timeline from work-level affiliation data in OpenAlex. Institutional relationships are resolved using ROR parent-child hierarchy and OpenAlex lineage data, producing four tiers: \texttt{SAME\_DEPT}, \texttt{SAME\_INSTITUTION}, \texttt{SAME\_PARENT\_ORG}, \texttt{DIFFERENT}.

When affiliation data is insufficient for classification, citations are labeled \texttt{UNKNOWN} (see Section~\ref{sec:unknown}).

\subsubsection{Phase 4: Venue Governance Detection}

This phase detects citations flowing through venues where the target researcher or their network holds governance roles. I use a locally deployed language model (Qwen 3.5 8B, Q4\_K\_M quantized) to extract structured governance data from venue editorial board pages.

\textbf{Stage 1 --- Database construction.} For each venue, the system fetches the editorial board page (via httpx or Cloudflare Crawl API for JS-rendered sites), feeds the raw HTML to the local LLM for structured extraction (member name, role, institution, ORCID), and performs entity resolution against OpenAlex author profiles. Results are cached with confidence scores and timestamps.

\textbf{Stage 2 --- Citation reclassification.} For each citation currently classified as \texttt{EXTERNAL}, I check whether the citing venue has governance overlap with the target researcher or their network. Classifications: \texttt{VENUE\_SELF\_GOVERNANCE}, \texttt{VENUE\_EDITOR\_COAUTHOR}, \texttt{VENUE\_EDITOR\_AFFIL}, or \texttt{VENUE\_COMMITTEE}.

The venue governance database is designed for incremental growth: each new analysis contributes new venues. A nightly job refreshes stale entries.

\subsubsection{Phase 5: Field Normalization and Percentiles (Proposed)}

Field-normalized percentile ranks comparing a researcher's scores against peer cohorts (same field, similar career length, comparable publication volume). This phase is not yet implemented.

\subsection{Author Identity Validation via ORCID}

A critical challenge I encountered during development was author disambiguation error in OpenAlex. Researchers with common names may have works misattributed to their profile. I address this through ORCID cross-validation using a smart two-mode system:

\begin{itemize}
    \item \textbf{High ORCID coverage ($\geq$70\%):} ORCID is used as a hard filter; only works in both ORCID and OpenAlex enter scoring.
    \item \textbf{Low ORCID coverage ($<$70\%):} All OpenAlex works are kept, but affiliation anomaly detection flags works from institutions never associated with the researcher.
\end{itemize}

If the publication span exceeds 25 years, a warning suggests using \texttt{--since YEAR} to set the career start explicitly.

\subsection{Transparency and Audit Trail}

Every run produces a timestamped JSON audit file documenting: the researcher profile, every work analyzed, every citation link, every classification decision with a human-readable rationale, the co-author graph, the affiliation timeline and institution hierarchy, data quality metrics, and the score computation.

Every output carries an inline disclaimer: ``BARON and HEROCON measure citation network structure, not research quality, impact, or integrity.''

\subsection{Technology Stack}

Python 3.11+ CLI application using Typer, Rich, and httpx. No database required for Phases 1--3. Phase 4 adds a persistent venue governance database (SQLite/PostgreSQL) and a locally deployed Qwen 3.5 8B language model (Q4\_K\_M quantized, served via llama.cpp on Kubernetes). A Gradio-based web interface provides no-code access for all Phase 1--3 functionality.

\textbf{Data sources (all free and open):} OpenAlex, 260M+ works, 100M+ authors, 2.8B+ citation links; ORCID Public API v3.0; ROR API v2; Cloudflare Crawl API (Phase 4 only).

\textbf{Performance.} For approximately 80 publications and 1500 citations, Phase 3 completes in under 90 seconds with approximately 100--150 OpenAlex API calls.

\subsection{The UNKNOWN Classification and Data Quality Reporting}\label{sec:unknown}

OpenAlex work-level affiliations are present for approximately 75\% of recent works, with lower coverage for older publications. When affiliation data is missing, the system cannot determine whether a citation is genuinely external or institutional.

Early versions defaulted such citations to \texttt{EXTERNAL}, creating systematic bias: researchers with poor metadata received artificially inflated BARON scores. This is both technically incorrect and ethically problematic.

I address this with \texttt{UNKNOWN}: citations are classified \texttt{UNKNOWN} when (a) the target researcher has no affiliation data for the relevant time period, or (b) no citing author has affiliation data for the citation year. \texttt{UNKNOWN} citations are excluded from both BARON and HEROCON calculations.

\textbf{Data quality metrics:} Classifiable citations count and percentage. Reliability rating: HIGH ($\geq$85\%), MODERATE ($\geq$70\%), LOW ($\geq$50\%), VERY LOW ($<$50\%).

I acknowledge that excluding \texttt{UNKNOWN} citations from the denominator creates a different form of selection bias. If \texttt{UNKNOWN} citations are systematically different from classifiable ones, disproportionately from developing countries with poor metadata, or from older publications, then the computed scores may not represent the true citation distribution. I identify a sensitivity analysis comparing different \texttt{UNKNOWN} imputation strategies as a priority for future work.

\section{Conceptual Foundations and Tool Landscape}\label{sec:related_works}

I position Citation-Constellation within the existing landscape by looking in two directions. The first is the conceptual literature that has established, over decades, that citation patterns are shaped by social network structure, that self-citation is a spectrum rather than a binary, and that metrics can distort the practices they aim to measure. The second is the practical tool landscape, the software that researchers actually use today to analyze citations, map networks, and evaluate scholarly output. The conceptual literature tells me why citation network decomposition matters. The tool landscape tells me that despite decades of theoretical insight, no existing instrument translates that insight into an individual-level, auditable diagnostic. This section maps both dimensions to make that gap precise.

\subsection{Conceptual Foundations}

\subsubsection{Citation-Based Research Impact Metrics}

The h-index \cite{hirsch2005index} remains the most widely recognized single-number measure of research impact. While elegant in its simplicity, the h-index has well-documented limitations: it is field-dependent, penalizes early-career researchers, and cannot distinguish between different types of citations \cite{waltman2012inconsistency, bornmann2005does, costas2007h}. Seglen \cite{seglen1997impact} made the classic argument that journal impact factors should not be used for evaluating individual research.

The g-index \cite{egghe2006theory} and field-weighted citation impact \cite{waltman2011towards} address some limitations but share the fundamental problem of treating all citations equally regardless of source. Waltman \cite{waltman2016review} provides a comprehensive review, concluding that existing approaches remain inadequate for capturing the multidimensional nature of research influence. Leydesdorff and Bornmann \cite{leydesdorff2011fractional} demonstrated how fractional counting of citations affects impact measurement across fields.

\subsubsection{Self-Citation Analysis}

Aksnes \cite{aksnes2003macro} demonstrated that self-citations constitute a significant fraction of total citations. Kacem et al. \cite{kacem2020tracking} provided a comprehensive analysis of self-citation patterns across disciplines. Critically, Ioannidis \cite{ioannidis2015generalized} introduced a generalized view of self-citation extending beyond direct author self-citation to include co-author self-citation, collaborative self-citation, and coercive induced self-citation. This taxonomy is directly relevant to my multi-layer approach. My BARON and HEROCON scores operationalize a similar decomposition, extending Ioannidis's conceptual framework into a computable, auditable system.

Fowler and Aksnes \cite{fowler2007does} demonstrated that self-citation increases subsequent citation from others. Seeber et al. \cite{seeber2019self} explicitly framed self-citations as strategic responses to the use of metrics for career decisions, directly connecting to my Goodhart's Law concerns (see Section \ref{sec:goodharts-law}).

\subsubsection{Network-Aware Citation Analysis}

\textbf{Invisible colleges.} Crane \cite{crane1972invisible} introduced informal networks of scientists who influence each other's work outside formal institutional structures. Zuccala \cite{zuccala2006modeling} developed a formal model reconciling structural and processual perspectives. These invisible colleges are precisely the "hidden networks" that my tool makes visible.

\textbf{Co-authorship and citation.} Newman \cite{newman2001structure, newman2004coauthorship} established foundational methods for analyzing scientific collaboration networks. Glänzel and Schubert \cite{glanzel2004analysing} found that collaborative papers receive more citations. Moody \cite{moody2004structure} traced increasing disciplinary cohesion. My Phase 2 co-author graph detection operationalizes these findings.

\textbf{Citation proximity.} Most directly relevant to my work, Wallace, Larivière, and Gingras \cite{wallace2012small} examined citation proximity using degrees of separation in co-author networks, finding that direct self-citations are relatively constant across fields (hovering around 20\% in natural and medical sciences, and 10\% in social sciences and humanities), while citations to co-author network members vary substantially. Their approach is conceptually identical to my BARON/HEROCON decomposition. My contribution extends their analysis with temporal affiliation matching, venue governance detection, ORCID identity validation, and a tool that any researcher can run on their own profile.

\textbf{Institutional and geographic effects.} Larivière and Gingras \cite{lariviere2010impact} demonstrated the Matthew Effect in citation. Wagner and Leydesdorff \cite{wagner2005network} analyzed international collaboration network structure. Pan, Kaski, and Fortunato \cite{pan2012world} uncovered the role of geography in shaping citation and collaboration networks.

\textbf{Citation homophily and diversity.} Hofstra et al. \cite{hofstra2020diversity} documented the diversity–innovation paradox: underrepresented groups innovate at higher rates but their contributions are discounted. This suggests that citation network structure may systematically disadvantage researchers less embedded in dominant citation communities. My UNKNOWN classification and data quality reporting attempt to make such biases visible.

\subsubsection{Citation Cartels and Gaming}

Baccini, De Nicolao, and Petrovich \cite{baccini2019citation} documented citation gaming at national scale. Edwards and Roy \cite{edwards2017academic} described perverse incentives. Smaldino and McElreath \cite{smaldino2016natural} modeled the evolutionary pressure for bad science driven by metric optimization. Meho \cite{meho2025gaming} found institutions with publication surges of up to 965\% driven by strategic optimization. These patterns illustrate Goodhart's Law as articulated by Fire and Guestrin \cite{fire2019over}.

My tool does not detect gaming \textit{per se}; rather, it detects network structure. The distinction matters because in-group citation is normal and often appropriate. Nevertheless, making network composition visible creates accountability infrastructure.

\subsubsection{Author Disambiguation and Identity}

Ferreira, Gonçalves, and Laender \cite{ferreira2012brief} surveyed automatic methods for disambiguation. OpenAlex \cite{priem2022openalex} uses machine learning. Schulz et al. \cite{schulz2014exploiting} exploited citation networks for large-scale disambiguation, validating my use of co-author graphs as identity evidence. ORCID \cite{haak2012orcid} provides researcher-maintained persistent identifiers. I use ORCID as a trust anchor for validating algorithmically assigned works.

\subsubsection{Responsible Research Assessment}

DORA \cite{dora2013} calls for moving away from single-number metrics. The Leiden Manifesto \cite{hicks2015bibliometrics} requires that quantitative evaluation support qualitative expert judgment. Wilsdon et al. \cite{wilsdon2015metrictide} provided a comprehensive UK policy framework. De Rijcke et al. \cite{rijcke2016evaluation} documented how indicator use distorts research behavior. Brembs \cite{brembs2018prestigious} demonstrated that prestigious journals struggle to reach even average reliability, which is relevant to my venue governance detection, as it separates the structural fact of editorial connection from any quality judgment.

\subsection{Existing Tools and Platforms}

The conceptual foundations reviewed above have been established for decades, yet the practical tool landscape reveals a striking absence. No existing instrument translates these insights into an individual-level citation network decomposition with auditable, per-citation classifications. I survey the most prominent tools below to make this gap concrete.

\subsubsection{Bibliometric Network Visualization}

\textbf{VOSviewer} (\href{https://www.vosviewer.com/}{vosviewer.com}; \cite{vaneck2010software}) is a widely used free tool for constructing and visualizing bibliometric networks, including co-authorship maps, citation coupling, co-citation clusters, and keyword co-occurrence networks. Developed at CWTS, Leiden University, it excels at mapping the structure of a research field at the macro level. VOSviewer visualizes \textit{relationships between entities} (authors, journals, institutions) but does not analyze the citation profile of a single researcher. It does not classify individual citations by network proximity, does not compute per-researcher scores, and does not produce audit trails. VOSviewer answers \textit{what does this field look like?}, whereas Citation-Constellation answers \textit{where do my citations come from?}

\textbf{CiteSpace} (\href{https://citespace.podia.com/}{citespace.podia.com}; \cite{chen2006citespace}) is a Java-based tool for visualizing and analyzing trends and patterns in scientific literature, with particular strength in detecting research fronts, burst detection, and temporal co-citation analysis. Like VOSviewer, CiteSpace operates at the field level rather than the individual researcher level. It maps how a body of literature evolves, not how a single researcher's citations decompose by network proximity.

\textbf{Bibliometrix / Biblioshiny} (\href{https://www.bibliometrix.org/}{bibliometrix.org}; \cite{aria2017bibliometrix}) is an R package (with a Shiny web interface called Biblioshiny) for comprehensive science mapping analysis. It supports co-authorship networks, keyword co-occurrence, thematic evolution, and three-field plots. Bibliometrix is powerful and flexible for field-level bibliometric analysis but, like VOSviewer and CiteSpace, it does not perform per-citation network proximity classification for an individual researcher. It can compute co-authorship statistics and self-citation rates but does not build a multi-layer classification system (self → co-author → institutional → venue governance → external) with graduated scoring and per-citation audit trails.

\subsubsection{Literature Discovery and Citation Exploration}

\textbf{Connected Papers} (\href{https://www.connectedpapers.com/}{connectedpapers.com}) is a web-based tool that builds a visual graph of papers related to a seed paper, using co-citation and bibliographic coupling similarity. It is designed for literature discovery (finding related work), not for analyzing an individual's citation profile. It shows how \textit{papers} are connected to each other, not how \textit{people} are connected to the researchers they cite.

\textbf{ResearchRabbit} (\href{https://www.researchrabbit.ai/}{researchrabbit.ai}) is a free tool that builds interactive visualizations of citation networks and co-authorship connections from a seed collection of papers. It excels at discovering related work and tracking research trends. Like Connected Papers, it is a discovery tool rather than a profile analysis tool. It does not decompose a researcher's incoming citations by network proximity or produce structural scores.

\textbf{Litmaps} (\href{https://www.litmaps.com/}{litmaps.com}) provides interactive citation maps showing how a set of papers are related through citation links, with timeline visualization and monitoring for new relevant publications. It is oriented toward systematic review and literature monitoring rather than individual citation profile analysis.

\subsubsection{Citation Context and Intent}

\textbf{scite.ai} (\href{https://scite.ai/}{scite.ai}; \cite{nicholson2021scite}) is the tool most conceptually adjacent to Citation-Constellation in its ambition to go beyond raw citation counts. Scite classifies citations by \textit{intent}, specifically whether a citation provides supporting, contrasting, or mentioning evidence, using deep learning on full-text citation contexts. This is a genuinely valuable dimension of citation analysis that addresses the question \textit{how is this work being cited?} Citation-Constellation addresses a different but complementary question: \textit{who is citing this work, and what is their network relationship to the author?} Scite tells you whether a citation supports or contradicts a finding; Citation-Constellation tells you whether a citation comes from a co-author, a departmental colleague, or an independent researcher. The two analyses are orthogonal and could, in principle, be combined. A citation from an independent researcher that provides contrasting evidence carries a very different epistemic signal than a supporting citation from a direct co-author. Scite requires institutional subscription for full access; Citation-Constellation is free and open.

\subsubsection{Proprietary Institutional Analytics}

\textbf{Scopus Author Analyzer} (\href{https://www.scopus.com/}{scopus.com}) provides author-level metrics including h-index, citation counts, document counts, and co-author lists. It can display self-citation statistics and field-weighted citation impact. However, it does not classify citations by co-authorship distance, institutional affiliation, or venue governance. It treats all non-self citations as equivalent. Access requires a Scopus institutional subscription.

\textbf{InCites} (\href{https://incites.clarivate.com/}{incites.clarivate.com}) is a Clarivate research evaluation platform built on Web of Science data that provides institution-level and author-level citation analytics, including percentile rankings, category-normalized citation impact, and collaboration metrics. InCites can identify international and institutional collaboration rates, but it does not decompose an individual researcher's incoming citations by the network proximity between citing and cited authors. It operates at the aggregate level (collaboration rates across a portfolio) rather than the per-citation level (classifying each individual citation). Access requires an institutional subscription.

\textbf{SciVal} (\href{https://www.scival.com/}{scival.com}) provides research performance benchmarking and analytics built on Scopus data, including collaboration metrics, topic prominence, and institutional comparison. Like InCites, it offers aggregate collaboration indicators but does not perform per-citation network proximity classification.

\subsubsection{Open Bibliometric Data Platforms}

\textbf{OpenAlex} (\href{https://openalex.org/}{openalex.org}; \cite{priem2022openalex}) is the open data platform that Citation-Constellation builds upon. OpenAlex provides the raw data (260M+ works, 100M+ authors, 2.8B+ citation links) but does not perform citation network decomposition. OpenAlex can report self-citation counts and provides the building blocks (author IDs, institutional affiliations, citation links) from which Citation-Constellation constructs its multi-layer classification. A researcher using OpenAlex directly would need to write custom code to replicate what Citation-Constellation automates.

\textbf{Semantic Scholar} (\href{https://www.semanticscholar.org/}{semanticscholar.org}) from the Allen Institute for AI provides a free academic search engine with author pages, citation graphs, and AI-powered paper recommendations. Its TLDR feature summarizes papers, and its citation context feature shows snippets of how papers are cited. Semantic Scholar does not perform network proximity analysis on a researcher's citation profile.

\textbf{Google Scholar} (\href{https://scholar.google.com}{scholar.google.com}) provides the most widely used free author profile with citation counts, h-index, and i10-index. It does not distinguish between self-citations and external citations, does not classify citations by network proximity, and provides no audit trail or transparency into how metrics are computed.

\begin{table}[H]
\centering
\resizebox{\textwidth}{!}{%
\begin{tabular}{llllll}
\toprule
\textbf{Tool} & \textbf{Primary Function} & \textbf{\shortstack[l]{Per-Citation\\Decomposition}} & \textbf{\shortstack[l]{Network\\Scoring}} & \textbf{\shortstack[l]{Free \&\\Open}} & \textbf{\shortstack[l]{Audit\\Trail}} \\
\midrule
\href{https://www.vosviewer.com/}{VOSviewer} & Field-level network visualization & No & No & Yes (free) & No \\
\href{https://citespace.podia.com/}{CiteSpace} & Research front detection \& trends & No & No & Yes (free) & No \\
\href{https://www.bibliometrix.org/}{Bibliometrix} & Comprehensive science mapping (R) & No & No & Yes (open-source) & No \\
\href{https://www.connectedpapers.com/}{Connected Papers} & Literature discovery from seed paper & No & No & Freemium & No \\
\href{https://www.researchrabbit.ai/}{ResearchRabbit} & Citation/co-authorship exploration & No & No & Yes (free) & No \\
\href{https://www.litmaps.com/}{Litmaps} & Citation timeline \& monitoring & No & No & Freemium & No \\
\href{https://scite.ai/}{Scite.ai} & Citation intent classification & Partially (intent, not network) & No & Subscription & No \\
\href{https://www.scopus.com/}{Scopus} & Author metrics \& self-citation & Self-citation only & No & Subscription & No \\
\href{https://incites.clarivate.com/}{InCites} & Institutional research evaluation & No (aggregate only) & No & Subscription & No \\
\href{https://www.scival.com/}{SciVal} & Research performance benchmarking & No (aggregate only) & No & Subscription & No \\
\href{https://openalex.org/}{OpenAlex} & Open bibliometric data platform & No (raw data only) & No & Yes (open) & No \\
\href{https://www.semanticscholar.org/}{Semantic Scholar} & AI-powered academic search & No & No & Yes (free) & No \\
\href{https://scholar.google.com/}{Google Scholar} & Author profiles \& citation counts & No & No & Yes (free) & No \\
\href{https://citation-constellation.serve.scilifelab.se}{\textbf{Citation-Constellation}} & \textbf{Per-citation network decomposition} & \textbf{Yes (multi-layer)} & \textbf{Yes (BARON + HEROCON)} & \textbf{Yes (free, open-source)} & \textbf{Yes (full JSON)} \\
\bottomrule
\end{tabular}%
}
\caption{Comparison of bibliometric tools.}
\label{tab:tool-comparison}
\end{table}

\subsection{Bridging the Gap}

The conceptual literature and the tool landscape summary (Table \ref{tab:tool-comparison}), taken together, reveal a specific and persistent gap. On one side, decades of research, from Crane's invisible colleges \cite{crane1972invisible} through Wallace et al.'s network-distance citation analysis \cite{wallace2012small} to Ioannidis's generalized self-citation taxonomy \cite{ioannidis2015generalized}, have established that citation patterns are shaped by social network proximity, that different network relationships mediate citation in structurally distinct ways, and that treating all citations as equivalent discards information that matters for understanding scholarly influence. This is settled science. The insight is not in dispute.

On the other side, the practical tools available to researchers in 2026 do not act on this insight at the individual level. VOSviewer, CiteSpace, and Bibliometrix visualize network structure beautifully, but at the field level, not the individual profile level. Scopus, InCites, and SciVal provide author-level metrics, but collapse all non-self citations into a single undifferentiated count. Scite.ai classifies individual citations, but by rhetorical intent (supporting, contrasting, mentioning), not by social network proximity. OpenAlex provides the raw data from which a network decomposition could be built, but building it requires custom engineering that few individual researchers can undertake. Google Scholar reports numbers without any decomposition at all.

Consequently, a researcher in 2026 who wants to understand the network composition of their own citation profile possesses no tool to answer that question. Such a researcher might want to know what fraction of their citations come from co-authors, from institutional colleagues, from editorial connections, or from genuinely independent researchers, yet no existing instrument provides this decomposition.

The knowledge exists in the literature. The data exists in OpenAlex. The computation is not prohibitively complex. However, the instrument does not exist.

Citation-Constellation is that instrument. It takes the conceptual framework established by Wallace et al. \cite{wallace2012small}, Ioannidis \cite{ioannidis2015generalized}, and others, operationalizes it through a multi-layer detection architecture with ORCID-validated identity resolution and temporal affiliation matching, packages it as both a command-line tool and a no-code web application accessible to any researcher with a browser, and documents every classification decision in a contestable audit trail. It does not invent the insight. Rather, it builds the tool that the insight has been waiting for. Furthermore, through the seamless integration of large language models and AI agents, it augments this robust architecture, bridging the gap between rigorous network analysis and universally accessible exploration.

I turn now to the implications, appropriate use cases, and structural tensions that Citation-Constellation and its BARON and HEROCON scores might create.

\section{Discussion}\label{sec:discussion}

\subsection{Framing: Structure, Not Quality}

BARON and HEROCON measure \textbf{citation network structure}, where in the social graph citations originate, not research quality, impact, or integrity. This distinction is the central design principle. A low BARON score might indicate a productive lab leader whose group naturally builds on their foundational contributions \cite{glanzel2004analysing}, a small-field researcher in a 50-person community where nearly everyone is a co-author's co-author \cite{newman2001structure}, or an insular practice that warrants self-examination. A high BARON score might indicate a cross-disciplinary thinker \cite{rafols2010diversity}, an early-career researcher who has not yet built a collaborative network, or a \textit{lone genius} working in isolation. The tool cannot distinguish these cases.

This reframing is not modesty. It is epistemic honesty about what the numbers can and cannot say. A BARON score of 40\% reports that 60\% of a researcher's citations come from within their detected network. Whether that reflects healthy collaboration, strategic citation behavior \cite{seeber2019self}, or simply the natural dynamics of a small field \cite{crane1972invisible} requires qualitative judgment that no metric can provide.

\subsection{Alignment with Responsible Research Assessment}

DORA \cite{dora2013} calls for moving beyond single-number metrics. The Leiden Manifesto \cite{hicks2015bibliometrics} requires that quantitative evaluation support qualitative judgment. De Rijcke et al. \cite{rijcke2016evaluation} documented how indicator use distorts research behavior, directly relevant to any new bibliometric tool, including this one.

I resolve the tension between introducing new quantitative metrics and the responsible assessment movement by positioning BARON and HEROCON as \textbf{supporting evidence for narrative evaluation}, not standalone metrics. The key distinction is between metrics designed to rank (h-index, impact factor) and metrics designed to describe (BARON, HEROCON). A ranking metric invites competition; a descriptive metric invites reflection. The Leiden Manifesto's first principle is satisfied when a metric provides structural context enriching human judgment rather than replacing it.

The audit trail makes this alignment concrete. A committee evaluating a researcher's portfolio could examine the audit file to understand \textit{why} a BARON score takes a particular value, whether from intensive collaboration, a small field, or editorial roles. The number alone is silent; the audit trail speaks. This approach resonates with Rafols et al.'s \cite{rafols2012journal} argument that responsible indicators should make complexity visible rather than hiding it behind a single number.

\subsection{The Goodhart Vulnerability}\label{sec:goodharts-law}

If BARON/HEROCON were adopted for evaluation, researchers would optimize for them, a direct manifestation of Goodhart's Law \cite{goodhart1984problems, fire2019over}. The strategies are predictable: soliciting citations from strangers, avoiding citation of relevant co-author work, publishing in unfamiliar venues. Edwards and Roy \cite{edwards2017academic} documented similar perverse incentives. Smaldino and McElreath \cite{smaldino2016natural} modeled the evolutionary dynamics.

A researcher could inflate BARON by (a) publishing under multiple name variants to split their co-author graph \cite{ferreira2012brief}, (b) rotating institutional affiliations, (c) establishing reciprocal citation arrangements outside their detected network \cite{baccini2019citation}, or (d) strategically managing ORCID records. Similar strategies have been documented for h-index manipulation \cite{seeber2019self, meho2025gaming}.

My mitigation: prominent disclaimers in every output, data quality reporting that makes scores visibly approximate, and positioning for self-reflection and policy research. I acknowledge these guardrails are imperfect; the history of bibliometric gaming suggests that sufficiently motivated actors will find ways to optimize any published metric \cite{fire2019over}.

\subsection{The Case for HEROCON as Experimental}

I do not have empirical evidence for the specific HEROCON weight values. As discussed in Section \ref{sec:hercon-weight-discussion}, these weights encode a formal hypothesis about the relative strength of different network pathways. The ordering is consistent with findings that co-authorship represents a stronger social tie than mere institutional proximity \cite{newman2004coauthorship, glanzel2004analysing}, and that transitive network connections carry meaningful but attenuated social influence \cite{granovetter1973strength, wallace2012small}. However, the cardinal distances between weights are not grounded in data.

Preliminary informal testing suggests that HEROCON scores are relatively stable under small weight changes for researchers with diverse citation profiles. However, scores can shift substantially for researchers whose citations are concentrated in a single category. This echoes a general challenge in composite indicators \cite{cherchye2006creating, greco2018methodological, oecd2008handbook}.

BARON, by contrast, is binary and requires no weight calibration. It is the methodologically robust contribution. HEROCON is a promising extension whose value depends on future empirical work, but even in its current form, the gap's existence (if not its precise magnitude) is robust to weight perturbation.

\subsection{Use Cases and Appropriate Deployment}

I identify three appropriate use cases, in order of decreasing confidence:

1. \textbf{Science policy research.} Analyzing citation network structure at the field level is the safest use case. Aggregate patterns are more robust than individual scores \cite{hicks2015bibliometrics}, and field-level analyses can reveal disciplinary norms \cite{wallace2012small} informing policy on evaluation practices and collaboration incentives. Comparing BARON distributions across career stages, institution types, or geographic regions could provide evidence for structural inequalities \cite{hofstra2020diversity, lariviere2010impact}.

2. \textbf{Self-reflection.} A researcher examining their own citation profile. The audit trail makes this actionable: inspect which co-authors cite most, whether institutional peers contribute substantially, and how external reach evolves over time. This aligns with calls for researcher self-awareness about citation practices \cite{aksnes2003macro, ioannidis2015generalized}.

3. \textbf{Contextualizing evaluation.} Providing structural context alongside other evaluation evidence, consistent with DORA \cite{dora2013} and the Leiden Manifesto \cite{hicks2015bibliometrics}. The structural context helps evaluators ask better questions; it does not answer them.

I explicitly discourage use in hiring, promotion, or funding decisions as a standalone criterion.

\section{Limitations}

\textbf{Coverage bias.} OpenAlex is English-heavy and recent-heavy \cite{zheng2025understanding, martin2021google}. Researchers in non-English-language traditions or with older publication records may have systematically lower coverage.

\textbf{Temporal resolution limits.} Affiliation data is derived from publication-time institutional records, not employment records. A researcher changing institutions mid-year may have citations misclassified \cite{donner2020comparing, aman2018does, robinsongarcia2019many}.

\textbf{The small-field problem.} In a 50-person community, nearly everyone may be a co-author's co-author at depth 2 \cite{newman2001structure}. BARON would classify most citations as in-group, reflecting field size rather than citation practice. I plan to address this through field normalization (see Section \ref{sec:future-work-field-normalization}).

\textbf{ORCID selection bias.} Researchers with complete ORCID records \cite{haak2012orcid} tend to be affiliated with well-resourced institutions \cite{porter2022orcid}. Consequently, my ORCID-based validation provides better protection for the researchers who perhaps need it least.

\textbf{Author disambiguation remains imperfect.} Despite ORCID cross-validation, researchers without ORCID records cannot benefit.

\textbf{HEROCON weights are not empirically calibrated.} Diverse profiles are robust to perturbation; concentrated profiles are sensitive (see Section \ref{sec:hercon-weight-discussion}). Until calibration is completed, HEROCON should be interpreted as indicative rather than definitive.

\textbf{UNKNOWN creates a conditioned sample.} If UNKNOWN citations are systematically different from classifiable ones, from developing countries with poor metadata \cite{rafols2015underreporting, zheng2025understanding}, older publications, or certain disciplines, then computed scores reflect a biased subset. I plan to address this through sensitivity analysis (see Section \ref{sec:future-work-sensitivity-analysis}).

\textbf{Department-level matching is noisy.} ROR lacks department-level identifiers for most institutions \cite{lammey2020solutions}.

\textbf{Venue governance database coverage.} Phase 4's incremental database means first-time analyses in underserved fields have limited coverage. Coverage improves with use, following a pattern common in incrementally constructed knowledge bases where initial sparsity gives way to density as more queries contribute data \cite{shin2015incremental, hofer2023construction}.

\textbf{LLM extraction accuracy.} Phase 4 relies on a locally deployed language model for structured extraction from heterogeneous HTML pages. Recent work has demonstrated that LLMs can perform structured information extraction from web pages with competitive accuracy, including zero-shot extraction without task-specific training \cite{arasu2003extracting, brinkmann2024llm}, but errors remain inevitable for non-standard page structures, non-English content, or dynamically rendered editorial boards.

\textbf{The tool detects correlation, not causation.} A citation classified as DIRECT\_COAUTHOR may represent a genuinely motivated citation, a co-author who cites the work because it is the best available reference, not because of social proximity. The classification identifies structural potential for network-mediated citation, not actuality. This distinction between structural position and actual influence is well-established in social network analysis \cite{borgatti2005centrality, bornmann2008whatdo}.

\textbf{The \textit{lone genius} problem.} The framework implicitly assumes that collaboration is the norm, an assumption that holds for most contemporary science \cite{wuchty2007teams} but not universally. A truly independent researcher who works alone would have a very high BARON score for reasons opposite to those intended. The tool cannot distinguish this from massive external reach. This reinforces that it is a structural, not a quality, metric.

\textbf{No empirical validation against ground truth.} I have not yet demonstrated that scores correspond to independently verifiable properties of citation behavior. Validation would require showing that scores differ between collaborative and independent researchers, that ORCID validation changes scores for common names, or that structural classifications align with self-reported citation motivations. The extensive literature on citing behavior \cite{bornmann2008whatdo, tahamtan2019whatdo} demonstrates that researchers cite for diverse scientific and non-scientific reasons, suggesting that structural classifications will capture meaningful but incomplete signal about citation motivation.

\begin{figure}[H]
   \centering
    \includegraphics[scale=.18]{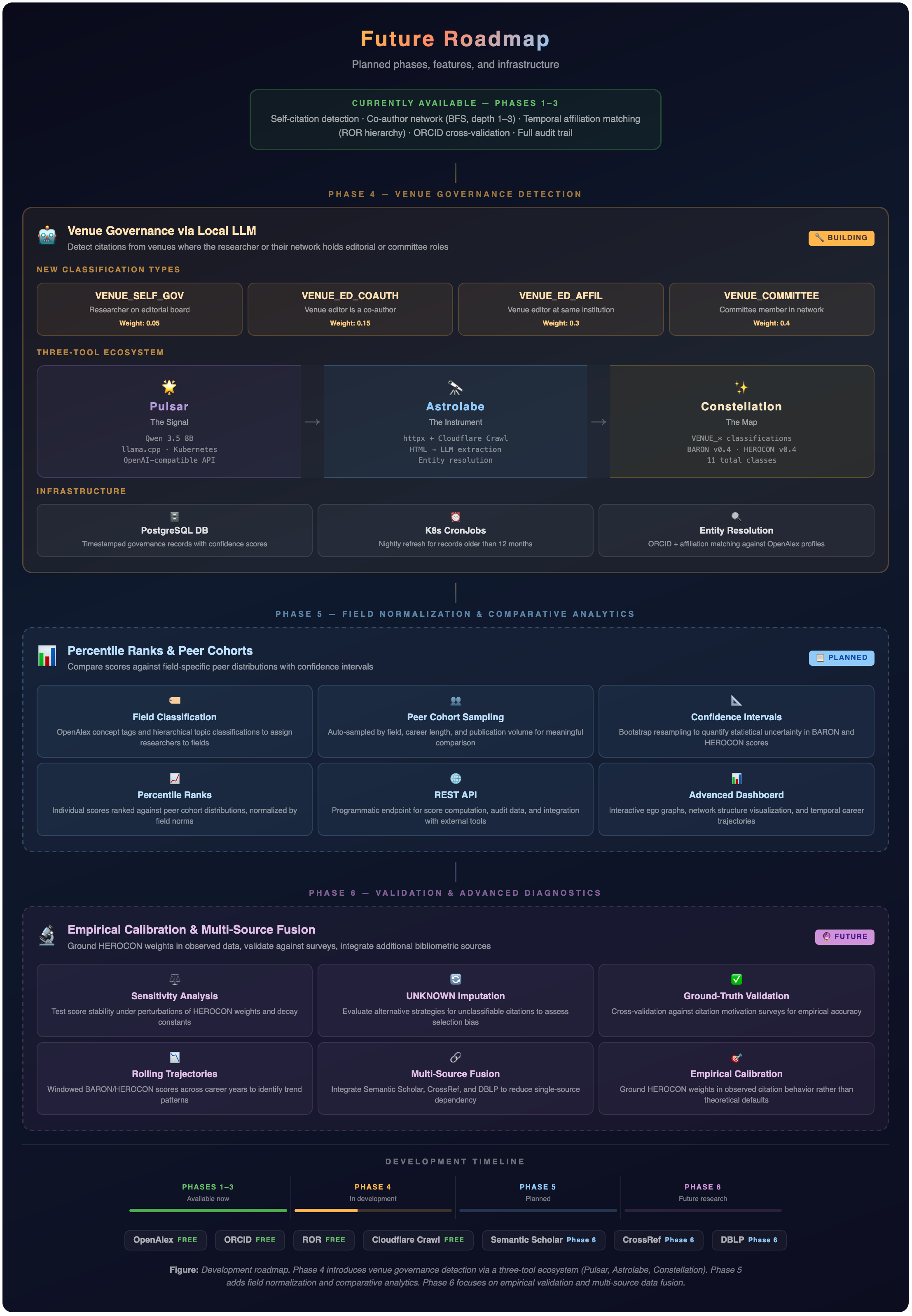}
    \caption{Development roadmap. Phase 4 introduces venue governance detection via a three-tool ecosystem (Pulsar, Astrolabe, Constellation). Phase 5 adds field normalization and comparative analytics. Phase 6 focuses on empirical validation and multi-source data fusion.}
    \label{future-contribution}
\end{figure}

\section{Future Work (Figure \ref{future-contribution})}\label{sec:future-work}

\subsection{Sensitivity Analysis (Priority)}\label{sec:future-work-sensitivity-analysis}

I plan to systematically examine how scores respond to perturbation along four axes: (a) HEROCON weight variation (±0.1, ±0.2) to map the sensitivity surface; (b) UNKNOWN imputation strategy (all-EXTERNAL, all-in-group, proportional allocation); (c) co-author graph depth (1 to 3); and (d) temporal decay rate (half-lives of 3.5, 7, and 14 years). For each axis, I will report score distributions across researchers from diverse fields and career stages.

\subsection{Empirical Weight Calibration}

I envision three approaches: survey-based citation motivation studies mapping self-reported motivations onto structural classifications; field-specific weight optimization using known collaborative versus independent profiles; and information-theoretic approaches maximizing discriminative power of the HEROCON–BARON gap. Calibrated weights would be published as field-specific weight files alongside the tool.

\subsection{Cross-Validation with Citation Motivation Studies}

Cross-referencing structural classifications against ground-truth data on citation motivation. I note that citation motivation is notoriously difficult to study; the goal is to establish whether the structural signal is informative, not whether it is deterministic.

\subsection{Field-Normalized Percentile Scoring (Phase 5)}\label{sec:future-work-field-normalization}

BARON 70\% in theoretical physics means something different than 70\% in biomedical research. Phase 5 will compute percentile ranks against peer cohorts, enabling statements like "BARON 72\% (65th percentile in computational biology)."

\subsection{Confidence Intervals}

Bootstrap confidence intervals reflecting sample size and metadata quality. A well-cited researcher with BARON 72\% ± 4\% presents a very different picture from an early-career researcher with BARON 72\% ± 27\%.

\subsection{Temporal Career Trajectory Analysis}

Windowed analysis (rolling 5-year BARON) to answer: \textit{Did external reach grow after changing institutions?} \textit{Did BARON decline after establishing a large lab?}

\subsection{Multi-Source Data Fusion}

Cross-referencing OpenAlex with Semantic Scholar, CrossRef, and DBLP \cite{zheng2025understanding} to improve coverage and assess score sensitivity to data source choice.

\subsection{Collaborative and Community Extensions}

Discipline-specific weight calibrations from domain experts, integration with institutional CRIS systems, support for non-Latin name scripts, and a shared venue governance database growing through distributed contribution.

\section{Conclusion}

In this paper, I have introduced BARON and HEROCON, two complementary bibliometric scores that decompose a researcher's citation profile into a structured map of self-citation, co-authorship, institutional proximity, venue governance, and genuine external reach. Where conventional metrics collapse all citations into a single number, these scores reveal \textit{where} in the social graph citations actually originate.

The contributions are both conceptual and architectural. Conceptually, the dual-score framework offers a new diagnostic lens. BARON's strict binary classification anchors the boundary of a citation profile by counting only fully external citations. HEROCON's graduated weighting maps the full constellation, assigning partial credit based on relationship proximity. The gap between the two serves as a readable diagnostic of inner-circle dependence, a quantity that, to my knowledge, no existing bibliometric tool reports.

Architecturally, I contribute an open-source, multi-layer detection system that combines self-citation analysis, co-authorship graph traversal with temporal decay, temporal affiliation matching via the Research Organization Registry, and AI-driven venue governance extraction into a single pipeline. The venue governance phase demonstrates a reusable pattern for AI-agent-driven bibliometric infrastructure. A locally deployed large language model performs structured extraction from heterogeneous web pages, an entity resolution layer matches extracted names against open bibliometric databases, and a persistent database captures the results, all running on commodity academic computing resources. This pattern is applicable well beyond venue governance and offers a template for automated extraction of any structured scholarly metadata that exists on the web but is not yet systematically captured.

Several engineering choices address persistent practical challenges that have historically prevented theoretical insights about network-mediated citation from becoming usable tools. ORCID-based identity validation mitigates author disambiguation error. The UNKNOWN classification honestly reports citations with insufficient metadata rather than silently inflating scores. Comprehensive audit transparency documents every classification decision with a human-readable rationale. And accessibility through both a command-line interface and a freely available no-code web application on SciLifeLab Serve ensures that citation profile analysis is not gated by technical skill, institutional subscription, or cost.

I am deliberate about what these scores are not. They are not quality indicators. They are not integrity detectors. They are not suitable for hiring, promotion, or funding decisions. They describe where in the social graph citations originate, nothing more, nothing less.

BARON is the methodologically robust contribution: binary, requiring no weight calibration, and directly interpretable. HEROCON is an experimental extension whose graduated weights encode testable hypotheses about the relative strength of different network pathways in mediating citation behavior. I have been transparent throughout this paper about what is established and what remains unvalidated.

If there is a single contribution I would ask readers to remember, it is the audit trail. Every classification decision is documented. Every uncertainty is flagged. Every assumption is exposed to scrutiny. In a field where metrics are too often treated as authoritative pronouncements, I offer a tool that shows its work and invites others to check it.

The tool is open-source and available. I welcome empirical evaluations, cross-field analyses, and community contributions toward grounding these structural diagnostics in evidence.

\begin{center}
\texttt{pulsar} $\rightarrow$ \texttt{astrolabe} $\rightarrow$ \texttt{constellation}\\[2pt]
{\small\textit{the signal \hspace{1.2em} the instrument \hspace{1.2em} the map}}
\end{center}

\begin{table}[H]
\centering
\small
\begin{tabular}{@{}lll@{}}
\toprule
\textbf{Component} & \textbf{Purpose} & \textbf{Link} \\
\midrule
Citation-Constellation
  & BARON \& HEROCON scoring
  & \href{https://citation-constellation.serve.scilifelab.se}{\textbf{No-Code Tool}} {\footnotesize|} \href{https://github.com/citation-cosmograph/citation-constellation}{Code} \\
Citation-Pulsar-Helm
  & LLM inference on Kubernetes
  & \href{https://github.com/citation-cosmograph/citation-pulsar-helm}{Code} \\
Citation-Astrolabe
  & Venue governance database
  & \href{https://github.com/citation-cosmograph/citation-astrolabe}{Code} \\
\bottomrule
\end{tabular}
\caption*{\href{https://github.com/citation-cosmograph}{The Citation-Cosmograph ecosystem on GitHub.}}
\end{table}

\section*{Acknowledgements}

This work would not exist without the infrastructure and generosity of others.

I am profoundly grateful to \textbf{SciLifeLab Data Centre} and the \textbf{SciLifeLab Serve} platform for providing the computational home where Citation-Constellation lives and breathes. Serve's commitment to hosting open research tools, freely, without barriers, for anyone, embodies exactly the democratic ethos this project aspires to. That a researcher anywhere in the world can navigate to a URL and decompose their citation profile without installing software, creating an account, or paying a fee is not my achievement alone; it is Serve's architecture that makes that sentence true.

I owe a particular and unusual debt to \textbf{Claude} (Anthropic). Citation-Constellation is, in the most literal sense, a project born in the quiet night hours after my daughter falls asleep, a pastime pursued after bedtime stories and lullabies. The idea of decomposing citation profiles by network proximity had lived in my head for some time, but the distance between an idea and an implementation is vast, and the hours between a child's sleep and one's own are few. Claude bridged that distance.

Finally, I thank my daughter, whose nightly surrender to sleep is the starting gun for everything I build in the dark. The constellations in this paper are named for heroes and boundary guardians, but the brightest star in my sky is considerably smaller. I hope that one day she will read this paper, understand very little, care even less, and still say, with the generous conviction that only a daughter can muster, "This is really cool! Abbu."

\section*{A Letter to the Reader}

If you have read this far, you have already seen what this tool does and how to use it. You may wonder why it exists, or why I built it alone, in the hours between my daughter's bedtime and my own.

It is now two years since my PhD. This paper represents my last formal scholarly work, not because I am leaving science, but because I want to leave something \textit{for} science that is not entangled with the incentive structures I have spent these pages critiquing. I built Citation-Constellation as a gift. It is not a stepping stone to tenure, not a citation magnet, and not a product. It is infrastructure that I needed and could not find, so I built it for us.

The idea came from a specific frustration: watching brilliant colleagues at under-resourced institutions struggle to demonstrate their impact to committees who only saw raw citation counts. Watching early-career researchers in small fields penalized for \textit{low} h-indices that reflected their community size, not their influence. Watching myself wonder, genuinely, whether my own citations came from intellectual reach or from the natural amplification of my immediate network. The tools to answer that question existed in theory, locked behind paywalls and technical barriers. I wanted to put them in a web browser, for free, with no registration required, because curiosity about your own scholarly footprint should not require an institutional subscription.

I am a new father, and I found that the only hours available for side projects are those quiet moments after lullabies. My daughter provided the structure: her bedtime ritual became the starting gun for every coding session. This tool carries the DNA of those nights. Imperfect, earnest, built under constraints, but offered with generosity.

I hope you use it to understand yourself better, not to rank yourself against others. I hope you use it to advocate for yourself in evaluation contexts, or to check your own biases in citation practices, or simply to satisfy a curiosity about the invisible architecture of your intellectual community. I hope you find the gap between BARON and HEROCON not as a judgment but as a map, showing where you stand in relation to the communities that sustain and cite you.

If this tool helps one researcher at a teaching-intensive college demonstrate their external reach to a promotion committee; if it helps one graduate student realize their citations are more diverse than they feared; if it helps one field recognize that their "low" citation counts reflect close collaboration rather than low impact, then those nights were worth it.

This is my gift to the scholarly community that sustained me. Use it kindly.

\begin{flushright}
Mahbub Ul Alam\\
Stockholm, Sweden\\
28 March, 2026
\end{flushright}

\bibliographystyle{unsrturl}
\bibliography{references}

\end{document}